\documentclass[twocolumn]{aastex62}
\usepackage{mathtools,float,hyperref}

\definecolor{darkgreen}{rgb}{0.09, 0.45, 0.27}

\defcitealias{CG97}{CG97}

\newcommand{\be}{\begin{eqnarray}}
\newcommand{\ee}{\end{eqnarray}}
\newcommand{\beqn}{\begin{eqnarray}}
\newcommand{\eeqn}{\end{eqnarray}}
\newcommand{\bi}{\begin{itemize}}
\newcommand{\ei}{\end{itemize}}

\newcommand{\dd}{{\partial}}

\def\refnew#1{(\ref{#1})}

\usepackage{ulem}
\newcommand{\x}{\sout}

\newcommand{\y}{\color{red}}
\newcommand{\w}{}
\newcommand{\bld}[1]{\mbox{\boldmath$#1$\unboldmath}}

\begin{document} 

\title{The Irradiation Instability of Protoplanetary Disks}

\author
{Yanqin Wu}
\affiliation{Department of Astronomy and Astrophysics, University of Toronto, Toronto, ON M5S 3H4, Canada}

\author{Yoram Lithwick}
\affiliation{Dept. of Physics and Astronomy, Northwestern University, 2145 Sheridan Rd., Evanston, IL 60208
\& Center for Interdisciplinary Exploration and Research in Astrophysics (CIERA)}
	
\begin{abstract}
The temperature 
in most parts of a protoplanetary disk is determined by irradiation from the central star.  Numerical experiments of \citet{Watanabe} suggested that 
such disks,  also called `passive disks', suffer from a thermal instability.  Here,  we 
use analytical and numerical tools to elucidate the nature of this instability. We find that it is related to the flaring of the optical surface,
 the layer at which starlight is intercepted by the disk.
 Whenever a disk annulus is perturbed thermally and acquires a larger scale height, disk flaring becomes steeper in the inner part, and flatter in the outer part.
Starlight  now shines more  overhead  for the  inner  part  and so can  penetrate  into  deeper  layers; conversely,  it 
 is  absorbed  more  shallowly in the outer part.  These  geometric changes allow  the  annulus  to  intercept  more  starlight, and the perturbation grows. We call this the irradiation instability. 
 It requires only ingredients known to exist in realistic disks, 
 and operates best in  parts that are {\w both} optically thick {\w and geometrically thin} (inside  $30$AU, but can extend to further reaches when, e.g., dust settling is considered).  An unstable disk develops 
travelling thermal waves that reach order-unity in amplitude. 
In thermal radiation, such a disk should
appear as a series of
 bright rings interleaved with dark shadowed gaps, while  in scattered light it
 resembles a moving staircase.
 Depending on the  gas and dust responses, this instability could  lead to a  wide range of consequences, such as {\w ALMA rings and gaps,} dust traps, vertical circulation, vortices and turbulence.

\end{abstract}

\section{Introduction}

 Currently, the main bottleneck for  understanding planet formation lies in an
 incomplete knowledge of the  protoplanetary disk.  In this work, we study the dynamics of passive disks,
i.e., disks where stellar irradiation
dominates the energetics. 
{\w These include almost all parts of the disk except perhaps for the innermost region.}
 We show that passive disks suffer from
 an instability,  and this could have  a broad range of  observational and theoretical ramifications.

\subsection{Motivation}

\noindent
{\bf I. Gaps and Rings:}
Recent observations made using the Atacama Large Millimetre Array (ALMA) have shown that typical disks are not the smooth power-laws beloved by theorists. Rather, bright rings and dark gaps are ubiquitous, on scales from  tens to hundreds of AU \citep[e.g.,][]{HLTau,Andrews,Huang19}. 
 At the moment, these
 rings and gaps are most commonly attributed to the effects of unseen planets
\citep[e.g.,][]{2014prpl.conf..667B,2015ApJ...809...93D,2016MNRAS.459L...1D,2017ApJ...850..201B,2017ApJ...843..127D,2018ApJ...869L..47Z}. 
And in a few cases, there is strong  kinematic evidence for planets, 
such as in  the gap of HD 163296 
	 \citep{Pinte2018}. 
But the near-ubiquity of gaps and rings \citep{2018ApJ...869...17L,2019AJ....158...13N}
is at tension with the paucity of large-mass planets  at these distances, as suggested by  direct imaging surveys \citep{Bowler_2016} 
and microlensing surveys \citep{2016ApJ...833..145S,2021arXiv210201715G}.
While
the planet hypothesis is difficult to exclude, given its large number of free parameters 
(such as planet mass and orbit, disk viscosity, and history)  it is worthwhile to consider whether planets are the causes for the gaps and rings, or instead the products of such features.

Many alternative scenarios 
 have   been proposed  to explain these features, including  dust-drift-driven viscous ring instability 
\citep{2005MNRAS.362..361W,2018A&A...609A..50D},
secular gravitational instabilities
in the dust \citep{Takahashi2014},
dead zones
\citep{2015A&A...574A..68F},
snow lines \citep{2016ApJ...821...82O},
MHD wind-driven structures 
\citep{2014ApJ...791..137B,2020A&A...639A..95R}, 
and an
eccentric disk instability \citep{2021arXiv210202216L}.
Lastly,
\cite{2012A&A...539A..20S} and \cite{2019ApJ...871...10U}
propose that they may be triggered by an 
instability
found in Monte Carlo simulations of irradiated disks.
This last proposal  may be
closely related to the irradiation instability considered
in this paper.

These features motivate us to study the stability of passive disks. An inherent instability could naturally explain the multiple gaps and rings in a given disk, without invoking an arbitrary number of planets. It would also directly impact the formation and migration of planets.

\bigskip
\noindent
{\bf II. Dust Wafting and Migration:} Radiation in disks is controlled by  dust.
Micron-sized grains absorb and scatter starlight, while
larger ($\sim$mm-sized) grains provide the bulk of the opacity for the thermalized radiation. 
Conversely, the distribution of dust grains is strongly influenced by gas dynamics.

 Both the spectral energy distribution  \citep[e.g.,][]{2001ApJ...547.1077C,Furlan,Woitke2019} and scattered light images \citep[e.g.,][]{2018ApJ...863...44A} of protoplanetary disks suggest that micron-sized grains must reside at least a couple gas scale heights above the midplane.
Although they are tightly coupled to the gas,  vertical settling high up in the atmosphere is rapid \citep{2004A&A...421.1075D}, even more so when dust coagulation occurs \citep{Dullemond05}. These small grains must be replenished, by fragmentation and/or up-draft. Both can be provided by gas turbulence or circulation.

The mm-sized dust, on the other hand, is much more weakly coupled to the gas, and so settles closer to the midplane. Without  any local pressure maxima, these grains can drift for large distances over the disk's lifetime \citep{1977MNRAS.180...57W}.  
Pressure maxima, on the other hand, can halt this migration and trap drifting particles
 \citep{1972fpp..conf..211W}.  ALMA images tantalizingly suggest that the observed rings and gaps are  the smoking gun for  particle traps \citep[e.g.,][]{2018haex.bookE.136A,Dullemond19}.

A disk instability could  therefore have important
implications for the dust  behaviour in protoplanetary disks. 
It could  generate turbulence or meridional flows that waft up the micron-sized grains.
And it may also produce pressure maxima, leading to vortices or axisymmetric rings,   natural barriers for inwardly drifting dust  and welcoming cradles for planet formation.

\bigskip

\noindent
{\bf III. Turbulence and Accretion:}
A crucial open question in the study of protoplanetary disks is why disks accrete \citep{LyndenBell1974}.  It is unclear if 
these disks are turbulent, and 
whether such turbulence
 can transport enough angular momentum to
disperse disks in a few million years
 \citep[see review by][]{2018haex.bookE.138K}.  Until recently, the most
promising mechanism for generating turbulence was the magnetorotational instability
\citep[][]{1998RvMP...70....1B}. 
But such MHD proposals require
a sufficient degree of ionization 
that the disk
gas can couple effectively to the magnetic field. That is hard to achieve,
particularly for disks that 
are very dusty \citep[e.g.,][]{2015ApJ...798...84B,2015MNRAS.454.1117S}.
A plethora of other instabilities have been investigated \citep[as reviewed in][]{2018haex.bookE.138K}.
One example is the vertical shear instability \citep{1998MNRAS.294..399U,2015ApJ...811...17L}, 
 which may produce sufficiently strong turbulence for accretion ($\alpha\sim 10^{-4}$, where
$\alpha$ is the Shakura-Sunyaev parameter), 
but at large distances
 \citep{2020ApJ...897..155F}.
 This $\alpha$-value appears consistent with some upper  limits placed at various disk locales 
\citep[e.g., $\alpha\lesssim 7\times 10^{-3}$,][]{2018ApJ...856..117F}.  
Alternatively, 
disks  may also disappear via a non-turbulent mechanism such as disk winds 
\citep[][]{2013ApJ...769...76B}. 

As the situation remains murky, we are motivated to look for a robust instability  that could generate turbulence or fluid circulation, and ultimately drive accretion.

\subsection{Prior Work on Passive Disks and Their Stability}

 For a protoplanetary disk accreting at a typical rate {\w (say, $\dot M \sim 10^{-8} M_\odot$/yr)}, 
disk heating is dominated by stellar irradiation except inward of $\sim 1$AU \citep[see, e.g.,][]{D'Alessio98}. So, most of the disk \x{is} {\w should be regarded as being} `passively heated.'
 \cite{KenyonHartmann} showed that  passive disks
 can account for  the far-IR excesses of IRAS disks,  provided they are flared. The predicted flaring morphology was
confirmed by  Hubble images \citep[e.g.,][]{1996ApJ...473..437B}.

    \citet[][hereafter CG97]{CG97} set forth   a simple model for such disks. 
Optical light from the star is absorbed
high up in the disk by small grains. And radiation from these grains illuminates the disk midplane. In thermal and hydrostatic equilibrium, flared passive disks take
 a simple analytic form: $\frac{h}{r}\propto r^{2/7}$, 
 where $h$ is the  vertical scale height and $r$ 
   the cylindrical radius
\citep{Kusaka,1976ApJ...208..534C,CG97}. 
Passive disk models have been very successful
in explaining 
the spectral energy distributions of protoplanetary disks, provided one also accounts
for some vertical settling of the dust 
  \citep[][]{2001ApJ...547.1077C,2004A&A...421.1075D,2006ApJ...638..314D}.   

The stability of passive disks  was   first investigated by \citet{1976ApJ...208..534C}  for disks around black holes. That work was extended to protoplanetary disks by \citet{D'Alessio99}.
Their simple analysis showed that  such disks are stable: thermal perturbations  propagate inwards, and damp  along the way.

 The equilibrium solution for a stable disk should be easily obtained by  iteration.
Yet mysteriously,  
such attempts  are often plagued by convergence issues
\citep[e.g.,][]{2004A&A...417..159D,2009A&A...497..155M,2012A&A...539A..20S,WangGoodman,2019ApJ...871...10U}.
Using Monte Carlo radiative transfer codes 
to describe the radiation effects more accurately than  \citetalias{CG97}, these authors 
iteratively solved the equations of hydrostatic equilibrium and thermal equilibrium.
They often find
 no convergence,  particularly for  disks with
 realistically large dust surface densities.   With successive iterations, new waves
 appear at large radii, and  propagate inward
 with order-unity amplitudes \citep[see, e.g., Fig.~7 of][]{2019ApJ...871...10U}.  It is unclear if such  behaviour is generic in physical disks, or if it is an artificial instability  introduced by the iteration procedure.
 In any case, this issue hampers further study of  realistic disks.

 The work by \citet{Watanabe}, though receiving little attention
 {\w \citep[see][for a modern reincarnation]{ueda}}, raises an interesting possibility. Using  1-D simplified radiative transfer, half-way in complexity between  \citetalias{CG97} and a Monte Carlo code, they found that passive disks are unstable.  They argued that the instability is likely related to changes in the optical surface. {\w Their numerical experiments showed that the unstable disk develops inward travelling thermal-waves that reach order unity amplitudes.}
 
In this work, we  further elucidate the origin of this behaviour,  which we call the `irradiation instability.' Using both analytical and numerical  tools (including the radiative transfer code RADMC-3D), we demonstrate that the instability is genuine,  not numerical -- although using RADMC-3D \citep[][henceforth called RADMC]{2012ascl.soft02015D} to iterate is risky unless guided by analytical insights. We derive the conditions for such an instability and argue that they   should 
be prevalent in observed disks.

\bigskip

{\bf Paper Overview:} 
In \S \ref{sec:mech} we present a cartoon view of the  instability.
In \S \ref{sec:staging}--\ref{sec:radmc} we do the math. 
Because of the complexities involved with solving the radiative transfer problem,
we  develop three models that are successively more complex, and more realistic. 
Readers who prefer to skip the technical details may proceed
to \S  \ref{sec:summary} where we summarize the main results. We end with an extensive discussion of the assumptions (\S \ref{sec:assumptions}) and a brief introduction of  things to come (\S \ref{sec:conclusion}).

\section{The Irradiation Instability --  Cartoon Version}
\label{sec:mech}

\subsection{Passive Disks in Equilibrium}

\begin{figure*}[!t]
    \includegraphics[width=0.47\textwidth]{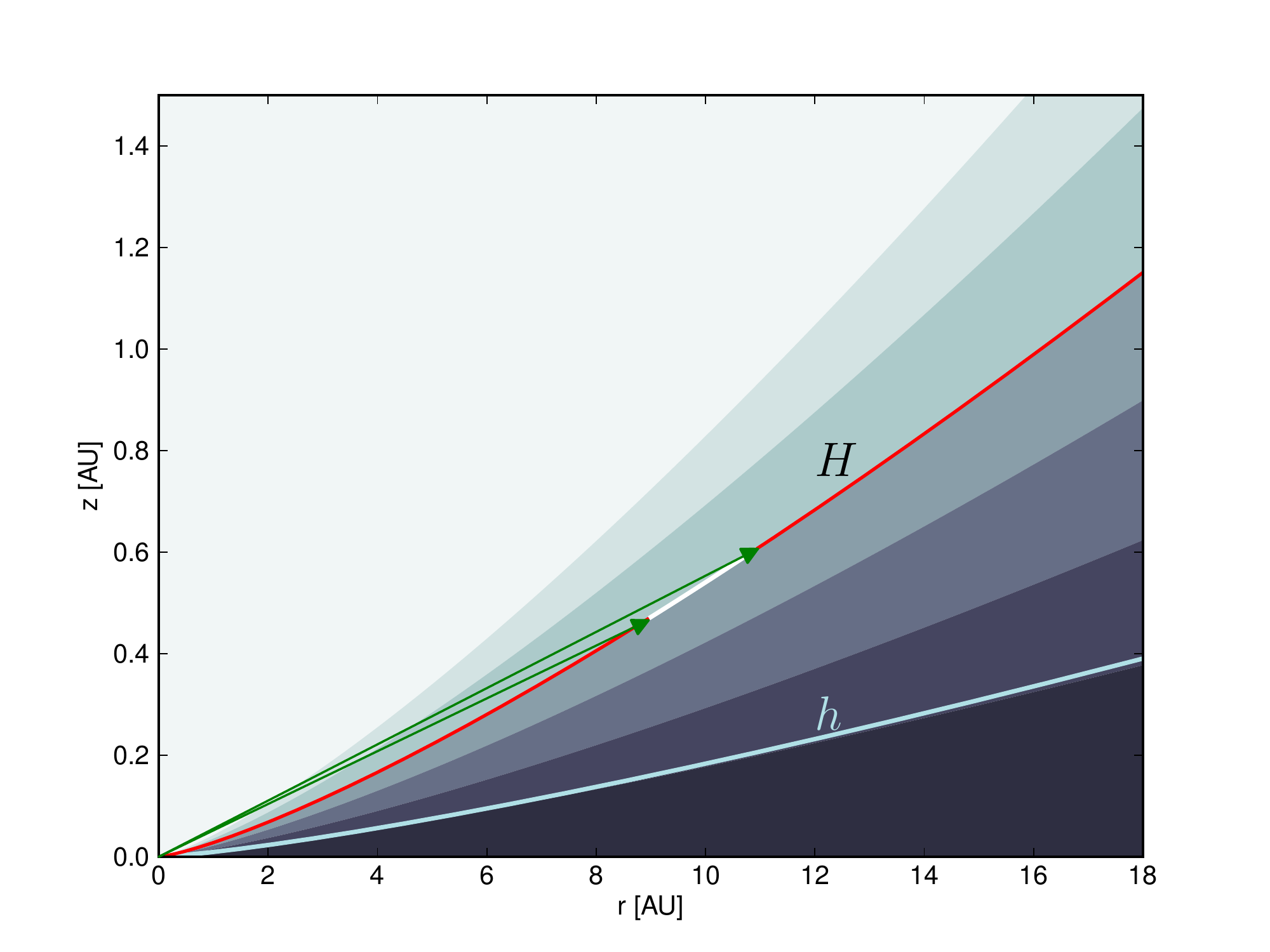}
    \includegraphics[width=0.47\textwidth]{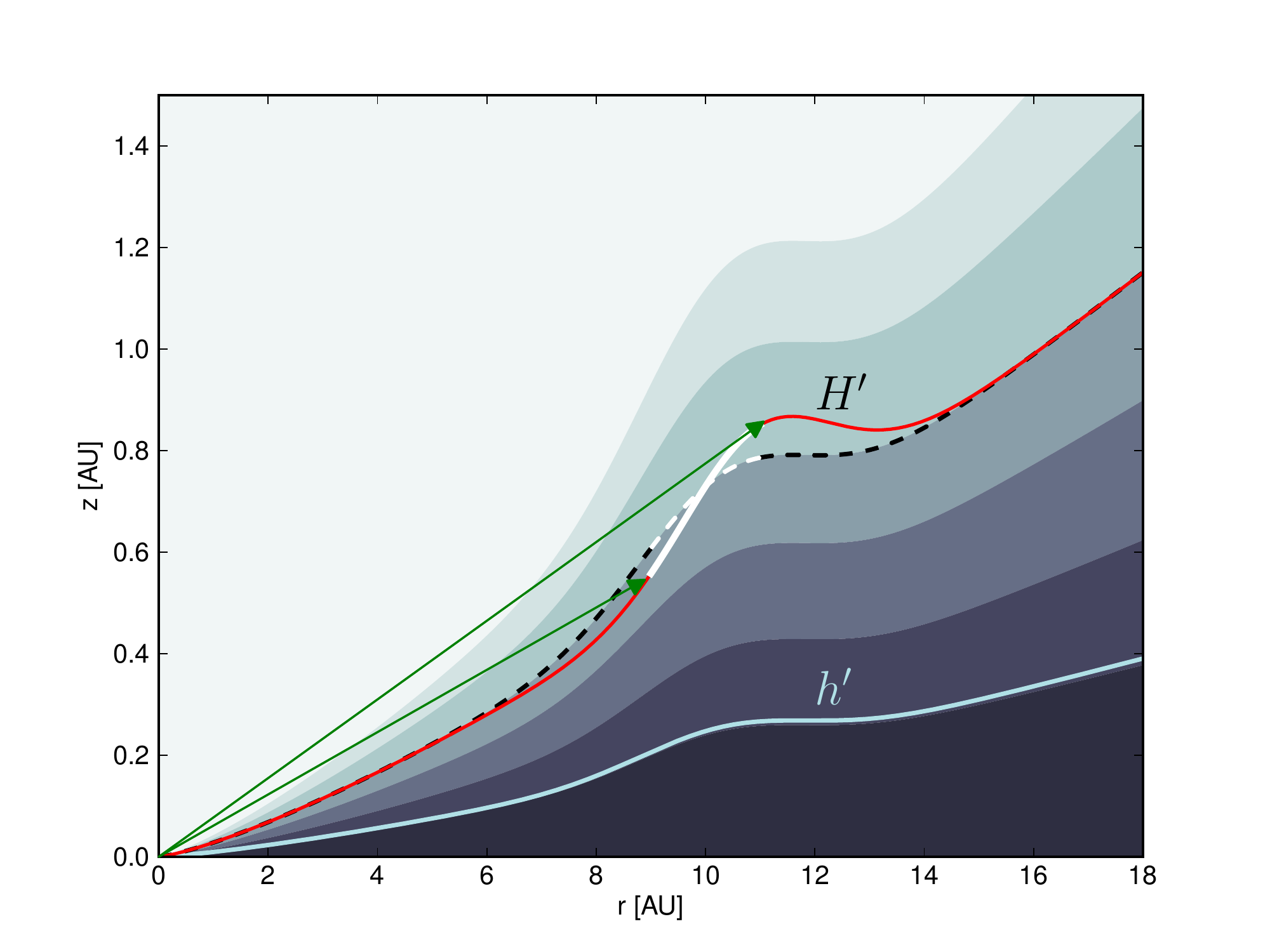}
    \centering
    \caption{Cartoon  of a passive  disk, in equilibrium (left) and perturbed (right).  The color shading illustrates the {\w dust density relative to that at midplane}.
    Here, $h$ is the vertical gas scale height, while $H$ (red line) is the  `optical surface,' the height at which radial starlight (green lines) is intercepted by the disk. The angular  extent of the white segment, as seen by the star, indicates the amount of starlight intercepted by the disk in that radial zone. 
      As this zone experiences
a rise in midplane temperature (right panel), 
the amount of starlight 
it receives increases, as evidenced by the larger angle subtended by the two green rays. This increase occurs for two reasons: the rise of the optical surface in proportion to the rise in scale height (dashed curve), and  changes in the penetration depths of the star rays. 
}
    \label{fig:unpert}
 \end{figure*}
 
We follow the concepts and notations
 of \citetalias{CG97} to study passive disks. 
 Fig.~ \ref{fig:unpert} (left panel) shows a cartoon of a passive disk in equilibrium, with the background shading representing the density of dust grains, relative to their midplane density. 
 The star's optical light is absorbed by  dust grains at altitude $H$, the  `optical surface,'  that lies a few scale heights  above the midplane. This layer  re-radiates 
 half
of the luminosity it receives down into the disk,  and the latter 
then re-emits this energy at longer wavelengths  (thermalized radiation).

The  amount of heating a disk  receives is determined by the flaring of  its optical surface. In Fig.~\ref{fig:unpert}, the relevant star-rays received by  a radial zone of concern  are those bound by the two green arrows. When balancing the heating against blackbody cooling, and insisting on vertical hydrostatic equilibrium, a 
flared solution is found for these disks \citep{Kusaka,CG97,Dullemond00}.

\subsection{Passive Disk Perturbed}

  Fig.~\ref{fig:unpert}  illustrates what happens to
  a disk when a localized thermal perturbation  increases the scale-height $h(r)$.
There are two effects. 
 First, the optical surface rises in proportion to the scale height (dashed curve).  This is what is considered in \citet{D'Alessio99}.
The disk intercepts a bit more stellar flux, but  not enough to overcome the extra blackbody  cooling  from a  now hotter disk. As a result, the perturbation damps away. 

But there is another effect.
Consider first
the inner  half of the scale height perturbation. The slope of the optical surface is  increased there,  so the star's rays shine more directly overhead (i.e.,  closer to the surface normal),
and  can penetrate more deeply  into the disk. This effect is  analogous to stellar limb-darkening, but  now for absorption.
Conversely, at the 
outer part of the perturbation, 
starlight penetrates more shallowly due to the more tangent slant. 
As a result, the opening angle between the two green rays
is increased. This means excess heating, and given the  right perturbation, it can overcome the excess cooling and drive an instability.

 Interestingly, an extreme manifestation  of  stronger flaring leading to enhanced heating is the inner rim of  a  protoplanetary disk \citep{Dullemond01}. In this region, the star's rays enter the disk almost head-on. The strong heating puffs up the inner rim into a wall.

In the following sections, we make this picture quantitative.

\section{Setting the Stage}\label{sec:staging}

To study  the  stability of a passively irradiated disk, we make a number of
simplifying assumptions, both to  highlight what we believe are the most relevant effects, and to facilitate a short  treatment. These include:

\medskip{}

\noindent -- opacity is provided only by dust;

\noindent -- dust traces gas with a constant ratio;
 
\noindent -- vertical hydrostatic equilibrium;

\noindent -- the radial profile of surface density  does not vary;

\noindent --  axisymmetry;

\noindent -- gas temperature tracks dust temperature;
 
\noindent -- star is point-like so all star rays are radial;
 
\noindent -- no inner hot rim, which  would otherwise cast a shadow.
 
\medskip

In addition to these, we also introduce a string of other simplifications, which we will describe as we go along. 
In \S \ref{sec:assumptions}, we assess \x{how} some of these assumptions {\w and how varying them} may qualitatively affect our results.

\subsection{The Thermal Equation}
\label{sec:eom}

We  adopt cylindrical co-ordinates ($r$ and $z$), and assume that the disk is sufficiently thin that we may work to leading order in $z/r$ (the small-angle approximation). 
The  midplane temperature $T$ is governed by the {\it thermal equation}
(obtained in Appendix A {\w under a number of assumptions}) \begin{equation}
\frac{3}{8}\, \tau_{\rm mm} c_p\Sigma_{\rm gas} \frac{\dd T}{\dd t} = 
{ \frac{1}{2} F_{\rm irr}[T]}
-\sigma_{\rm SB}T^4    \ . \label{eq:thermal0}
\end{equation}

This describes the rate of thermal energy increase per unit disk area 
as a result of the imbalance between starlight heating 
and black-body cooling. 
Here, $c_p$ is the specific heat per unit mass, $\Sigma_{\rm  gas}$ is the gas  surface density, and $\tau_{\rm mm}$ is the vertical optical depth  for thermal radiation (more details below), which is assumed to be larger than unity.
On the right-hand side, $F_{\rm irr}$ 
is the stellar flux incident on the disk's optical surface.
The square brackets denote that $F_{\rm irr}$ depends on the local profile (not just value) of temperature.
And the factor of $1/2$ multiplying $F_{\rm irr}$ arises because half of  the incident flux is re-radiated by grains in this surface to space, without
heating the disk interior.

In writing eq. (1), we have simplified the physics of vertical
transport under the assumption that  the timescale of variation is much longer than the thermal time (see Appendix A). A more accurate treatment should allow for vertical thermal waves. We argue in Appendix \ref{subsec:A3} that this likely does not impact our results.

As we show in Appendix \ref{sec:thermaleq},  the $T$ that appears throughout equation \refnew{eq:thermal0} should be the midplane temperature, not the surface temperature, provided the factor of  $\tau_{\rm mm}$ is included on the left-hand side.\footnote{
The factor of $\tau_{\rm mm}$ has been incorrectly neglected in the literature \citep{D'Alessio99, Dullemond00, Watanabe}. But its neglect leads only to a  change in timescale.
}

 The heating flux is
 \citep{1962SvA.....6..217S,Kusaka,CG97}
  \be
F_{\rm irr} = \frac{L_*}{4 \pi r^2} \, \sin\alpha \, ,
\nonumber
 \ee
 where $\alpha$ is the  grazing angle 
of the star-rays relative to the  optical surface.
 We denote the height of this surface above the midplane by $H$.
In the limit that the  disk is thin ($H/r\ll 1$) and the star is point-like,
 we can simplify the expression for $\alpha$ to arrive at  \citep[\citetalias{CG97},][]{2001ApJ...547.1077C}, 
\be
F_{\rm irr} {\approx}  
\frac{L_*}{4\pi r^2} \, \left( \frac{d}{d\ln r} \, \frac{H}{r}\right) \ .
\label{eq:hthin}
 \ee
If $H/r$ decreases with radius, 
the disk falls into shadow  cast by
 interior annuli, and one should instead set $F_{\rm irr}\rightarrow 0$. 
 {\w Hidden in this equation is the implicit assumption that the stellar heating is only determined by the local gradient of the optical surface. This is reasonable as long as the variation length scale is much greater than $H$, as is certainly satisfied by the equilibrium disk. However, this may not be true for a perturbed disk, an issue we call `horizontal averaging' and return to later. }

In hydrostatic equilibrium, the midplane temperature 
determines the gas scale height as 
\be
\frac{h}{r} = \frac{c_s}{\Omega r}   
=  \left(\frac{k_B}{\mu m_p GM_*}\right)^{1/2} 
 \sqrt{rT} \ , \label{eq:hbyr} 
 \ee
 where $M_*$ is the stellar mass, $m_p$  the proton mass, and $\mu$  the mean molecular weight (henceforth set to $\mu=2.3$).
The scale height in turn
 sets the  dust density field, which we take to be 
 \begin{equation}
 \rho_{\rm  dust} (r,z)=\frac{\Sigma_{\rm  dust}(r)} {\sqrt{2\pi}h(r)}\, \exp\left[-\frac{z^2}{2h(r)^2}\right]\, ,
 \label{eq:rho}
 \end{equation}
  where $\Sigma_{\rm dust}$ is the vertical column density of dust (both sides of equator).
The  dust density controls how far the stellar flux  penetrates into the disk. 

In this study, we assume  that  vertical hydrostatic equilibrium is maintained instantaneously, or eqs. \refnew{eq:hbyr}-\refnew{eq:rho} remain valid as the disk heats and cools. This is  likely  valid in the  region where the thermal time is  longer than the orbital time,  or inwards of $\sim 50$AU  for our fiducial disk (see below).

\subsection{The Optical Surface}
\label{sec:twoa}

The  key  ingredient 
in this problem is the
geometry of the optical surface.   For our analytical study  (not the RADMC simulations), we
consider two approximate
forms for $H$:
\begin{enumerate}
	\item  Simplistic Surface:
	  following \citetalias{CG97}, we introduce an important quantity $\chi$, the ratio between the optical surface height and the local gas scale height,
		\be
	  \chi\equiv {H\over h}\ .
	  \label{eq:chi}
	\ee 
 The value of $\chi$, for passive disks in equilibrium, depends on dust density logarithmically (see more below). \citetalias{CG97} found that the value of $\chi$ ranged from $5$ to $4$ between  3AU and 100AU, for a MMSN-type disk. 
So at first sight it seems reasonable to  assume that $\chi$ is a constant everywhere, and does not vary when the disk is perturbed. We call this a `simplistic surface.' It is  adopted by \citet{D'Alessio99} for their stability analysis. 

	\item 
Realistic Surface: alternatively, one could 
self-consistently determine the optical surface using the definition that  the optical depth to the star is unity. 
 This brings in some algebraic difficulties but, as we show below, is essential for describing perturbed disks.  The name `realistic' is a euphemism -- this approach remains an approximation to reality, which can be addressed only using radiative transfer codes.
	
\end{enumerate}

In \S \ref{sec:constchi}-\ref{sec:unst}, we  study passive disks  under  each  of these approximations.  The key result is that under the first approximation the disk is stable, whereas under the second more accurate one
it is unstable. We then perform RADMC simulations (\S \ref{sec:radmc}) to demonstrate that true disks
also exhibit instability. 

\subsection{Fiducial Disk}
\label{sec:fiducial}

 We assume the star has  solar mass and  luminosity, {\w $M_* = M_\odot$, $L_*= L_\odot$}.
For our fiducial disk, we choose a dust surface density 
\be 
\Sigma_{\rm dust} &=& 20\,  r_{\rm AU}^{-1} \ \  \textrm{g}/{\rm cm}^2\, 
\label{eq:sigmadust}
\ee
 where $r_{\rm AU} = (r/{\rm 1AU})$.
This density is similar to the
minimum mass solar nebula at 1AU 
\citep{1981PThPS..70...35H}, but falls off more gradually  with radius, 
and so is  more consistent with many observed systems 
\citep[e.g.,][]{2016ApJ...832..110C,2017ApJ...837..132V}.

 We adopt the dust opacity from  Fig.~2  of \cite{2016A&A...586A.103W}, calculated for a power-law dust mixture with sizes from $0.05\mu m$ to $3$mm,  which we approximate as
  \begin{eqnarray}
\kappa_\lambda &=& 10^3\left({\lambda\over 0.5\mu m}  \right)^{-1/2} {\rm cm}^2/\textrm{g(dust)}  \ ,
\label{eq:kap}
\end{eqnarray}
for total extinction. 
In the optical, this opacity falls below  that adopted by \citetalias{CG97} by a factor of 40, as they assume all grains are  small
($0.1\mu$m).

\begin{figure}
	\begin{center}
	\includegraphics[width=1\columnwidth,trim=0 150 0 100,clip]{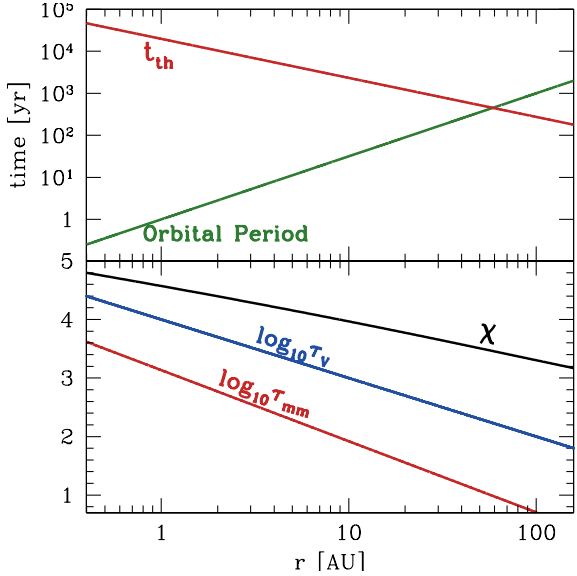}
        \caption{ Our fiducial disk.
        The top panel  compares the thermal time (red curve) and the
  dynamical time (orbital period, green curve); inward of $\sim 50$AU,  it may be reasonable to assume that vertical hydrostatic equilibrium is reached quickly. The bottom panel displays the vertical optical depths for visual (blue) and for thermal radiation (red). The disk is optically thick throughout.
    Here, the disk temperature profile is that of the equilibrium disk, and $\chi = {H\over h}$, both determined in  \S \ref{sec:phys}. 
       } 
    \label{fig:timescales}
    \end{center}
\end{figure}

In our analytical study, the radiation field is described by fluxes at only two frequencies: that of starlight and of disk thermal radiation. So
 the disk radiative properties can be  encapsulated by two vertical optical depths. One is $\tau_V$, the optical depth in the visual band,
 \be
 \tau_{\rm V} \equiv {1\over 2} \, {{\kappa_V  \Sigma_{\rm dust}}  }= 10^4\, r_{\rm AU}^{-1} \ , \label{eq:taufid}
 \ee
  where the factor of $1/2$ indicates integration from the midplane upward.
 The other is the optical depth for  dust thermal radiation. We name it  $\tau_{\rm mm}$,  with the actual wavelength  determined by the peak of the local blackbody. 
Fig.~\ref{fig:timescales} (bottom panel) shows these two optical
depths.
Our fiducial disk remains optically thick  to  thermal radiation out to $\sim 100$AU.

 Equation \refnew{eq:thermal0} induces us to define an evolutionary timescale, the thermal time, 
\begin{eqnarray}
  t_{\rm th}={3\over 8} {c_p\, \Sigma_{\rm  gas}\, \tau_{\rm mm}\over\sigma_{\rm SB}T^3}\, .
  \label{eq:t_thermal}
\end{eqnarray}
 We evaluate this timescale using  the equilibrium disk temperature, and a gas-to-dust ratio $\Sigma_{\rm gas}/\Sigma_{\rm dust}=100$,  and plot it in Fig.~\ref{fig:timescales}. It decays outward  as $t_{\rm th} \propto r^{-0.93}$, and intersects the disk dynamical timescale at around $50$AU.  Outside this radius, the assumption of vertical hydrostatic equilibrium fails. 

\section{Stability for  Simplistic Surface}
\label{sec:constchi}

We solve the thermal equation under the assumption that 
$\chi=H/h$ is a constant  in both space and time. While allowing for a spatially varying $\chi$ does not much alter our conclusions, a time-varying $\chi$ will, as we discuss later. 
We will present the equilibrium and a simple stability  analysis, 
neither of which is
new to this work. {\w We do so  both to  connect to the previous (mis)-understanding that passive disks are stable, and to highlight the differences between such a treatment and a more correct one (\S \ref{sec:unst}).}

\subsection{Equilibrium Disk} 
\label{sec:eqc}

The equilibrium solution has been presented by 
\citet{Kusaka}, \citetalias{CG97}, and \citet{Dullemond00}.
The thermal equation reads
 \be
{L_*\over 8\pi\sigma_{\rm SB}}{d\over d\ln r}{H\over r}=r^2T^4 \ . \label{eq:bg}
\ee
It may be integrated after setting $H=\chi h$ . With $\chi$  being a  constant, and $h$ determined by  hydrostatic equilibrium (eq.~\ref{eq:hbyr}), this equation has  a power-law
solution,\footnote{
The general solution for $h/r$ is
 given by eq.~\ref{eq:th2ans} multiplied by the factor $ \left(1+ a r^2\right)^{-1/7}$,
where $a$ is
an integration constant
\citep{Dullemond00}.
Here we focus only on the power-law part, as it is relevant for most of the disk.}
\be
{h\over r}
=
0.02\, \,  \chi^{1/7}\,  r_{\rm AU}^{2/7} \times \left( {L_*\over L_\odot}\right)^{1/7}\times \left({M_\odot\over  M_*} \right)^{4/7}
 \  , \label{eq:th2ans}
\ee
and
\be
T 
= 90 {\rm K} \,  \chi^{2/7}\, r_{\rm AU}^{-3/7}\times \left( {L_*\over L_\odot}\right)^{2/7} \times \left({M_\odot\over M_*} \right)^{1/7} \ .
\label{tempans}
\ee
These power-laws reproduce  eqs.~(14) in \citetalias{CG97}, albeit with slightly different normalizations.

{\w
As described below (\S \ref{sec:phys}), $\chi$ can be solved for self-consistently. It varies slowly for a disk in equilibrium, so eqs. (\ref{eq:th2ans})--(\ref{tempans}) remain approximately correct. 
In particular, for our fiducial disk, $\chi$ may be adequately fitted as  $\chi\approx 4.5 r_{\rm AU}^{-.075}$, and we estimate the optical surface to lie at 
 \begin{equation}
 {H\over r} = \, \chi {h\over r} \, \approx 0.11 \, r_{\rm AU}^{0.21}\, .
    \label{eq:scalingHold}
\end{equation}
}

\subsection{Linear Perturbations }
\label{sec:lincc}
We repeat  the  stability analysis by
\citet{1976ApJ...208..534C} and
\citet{D'Alessio98}, but using a form that will make generalization  to a more complicated form of $F_{\rm irr}$ straightforward.
We  perturb by setting
    $T\rightarrow T+\delta  T$,  and  similarly $F_{\rm irr}\rightarrow F_{\rm irr}+\delta F_{\rm irr}$. 
Linearizing the thermal equation yields
\be
{\dd\over \dd t}{\delta T\over T}={1\over t_{\rm th}}
\left({\delta F_{\rm irr}\over {F_{\rm irr}}} -4{\delta T\over T}  \right) \ .
\label{eq:lin}
\ee
The power-law equilibrium profile,  
 together with a constant $\chi$, yield
 \begin{eqnarray}
{{\delta F_{\rm irr}}\over{F_{\rm irr}} } & = &
 \left({ 
    {d \over{d \ln r}} {H\over r}}\right)^{-1} \, 
    { {\partial \over{\partial \ln r}}\left( {{\delta H}\over H} {H\over r}\right)}
    \nonumber \\
& = &    {7\over 4} 
{\partial \over{\partial \ln r}} {{\delta T}\over T} + {1\over 2} {{\delta T}\over T}\, ,
    \label{eq:added}
\end{eqnarray} 
 where the first term on the right-hand side bestows the travelling wave nature  for the perturbation, while the second term  provides positive feedback.
Now the thermal equation becomes
\be
 {\dd \over \dd t}{\delta T\over T} ={1\over t_{\rm th}}
  \left(  
 {7\over 4}{\dd\over \dd\ln  r}{\delta T\over T}-{7\over 2}{\delta T\over T} \right)\, .
 \label{eq:simpleq}
\ee
This equation admits a 
decaying travelling wave solution. 
We consider a complex $\delta T$ of the form
\be
{{\delta T}\over T} \propto e^{s t +   ik\ln r} \ , \label{eq:complx}
\ee
with  wavenumber $k$ (a real number, not restricted to integers) and  growth-rate $s$ (complex-valued).  The physical perturbations are understood to be the real parts  of these.
Inserting this into Equation (\ref{eq:simpleq}) then yields the growth rate 
 \be
 s  = {1\over t_{\rm th}}\left( {7\over 4}ik-{7\over 2}  \right)\, , \label{eq:sold}
 \ee
or, the thermal perturbation propagates inwards
with a  phase speed   ${d\ln r\over dt}\vert_{\rm phase}=-{1\over k}{\rm Im}[s]=-{7\over 4}{1\over t_{\rm th}}$, and decays at a rate 
${\rm Re}[s]=-{7\over 2}{1\over t_{\rm th}}$ \citep{D'Alessio98}.

Why is the disk stable?
 Naively, one might imagine that a hotter region rises up in  scale height and can therefore intercept more stellar flux. However, 
under the assumption that $\chi$ is constant, stellar heating scales with temperature to the half power ($F_{\rm irr} \propto  H \propto h \propto T^{1/2}$), while cooling scales with a higher power ($\propto T^4$). 
So the perturbation decays in time.

To understand why perturbations propagate
inwards, 
we consider the disk surface in the presence of a local  positive temperature perturbation. The inner half of the
affected region 
has a steeper grazing
 angle than in  equilibrium, and so is heated more. This raises the local surface under hydrostatic adjustment.
 Conversely, in the outer half, the surface drops. 
As a result, the perturbation moves inwards.

\section{Instability for Realistic Surface}
\label{sec:unst}

A crucial effect not accounted for in  the simplistic surface model is that the depth of starlight penetration changes when the slope of the  optical surface  varies (Fig.~\ref{fig:unpert}). That strengthens the heating response and leads to an instability.

Here, we  locate the disk surface  using the definition that the optical depth to the star is unity. 
The optical depth along a slanted ray from the star that has an inclination angle  $\theta_H$ is \citep{Watanabe}.
\be
\tau_{\rm  slant}&=&
\left( 1 + \tan^2\theta_H  \right)^{1/2}
\int_0^{r} \kappa_V\rho_{\rm dust}(r',z=\tan\theta_H r') dr'
\nonumber \\
&\approx&
\int_0^{r} \kappa_V\rho_{\rm dust}(r',z=\theta_Hr') dr' \ ,
\ee
where the approximate sign holds when $\theta_H\ll 1$. 
 With the  density profile given by eq.~\refnew{eq:rho}, and the vertical optical depth $\tau_V = \kappa_V \Sigma_{\rm dust}/2$, we have 
 the following equation to  determine the more realistic  optical surface, 
\be
1 = \tau_{\rm slant} \approx \int_0^r  dr'{2\tau_V(r')\over\sqrt{2\pi}h(r')}e^{- {1\over 2}{H(r)^2\over r^2}{r'^2\over  h(r')^2}}\, . 
\label{eq:phot}
\ee
 This is  our revised
 equation for the optical surface. Given a disk opacity and temperature profile, one can use it to solve for $H(r)$.
 
\subsection{An  approximate form and  qualitative discussion}
\label{sec:phys}

 To better understand 
the implications of eq.~\refnew{eq:phot}, we derive  a  simpler form that is valid  for $\chi\equiv {H\over h}\gtrsim$ a few.
  The optical depth  along a given line of sight ($\theta_H$) is mostly produced by material close to the optical surface, so we can write  $\tau_{\rm slant}
 \approx \rho_{\rm  dust}(r,H)\, \kappa_V\, \Delta r$, where  
 $\Delta r$ is  the  local density scale height as experienced by the slanted ray. Expressing density as  $\rho_{\rm  dust} (r,H)  \propto e^{-H^2/(2h^2)} \propto 
  e^{-r^2\theta_H^2/(2h^2)}$   (eq.~\ref{eq:rho}), 
  we have 
   \be
     {\Delta r}=\left.\left({d\ln\rho_{\rm  dust}\over d r}\right|_{\rm constant \,  \theta_H}\right)^{-1}={r\over \chi^2\gamma} \ ,
     \label{eq:deltar}
  \ee
  after defining a flaring index $\gamma$,
 \be 
 \gamma\equiv {d\ln (h/r)\over d\ln r} \ .
 \label{eq:gammadef}
 \ee 
 This index equals $2/7$ for the power-law disk (eq.~\ref{eq:th2ans}).
 Setting $\tau_{\rm slant} = 1$ then yields our 
desired approximate form for the optical surface,\footnote{Equation \refnew{eq:ape} may also 
be derived directly from equation \refnew{eq:phot}
as follows: assuming $\chi$ is large, the integrand in equation \refnew{eq:phot} is 
exponentially suppressed unless $r'$ is close to $r$. 
Therefore we may approximate $r'\approx r$ outside of the exponential, and inside of the exponential we may set
${r'^2\over h(r')^2}\approx {r^2\over h(r)^2}(1-2\gamma{r'-r\over r})$, in which case the  integral yields equation \refnew{eq:ape} \citep{2007ApJ...654..606G}.
\label{foot:garaudapprox}}
 \be
{\chi^2\over 2}e^{\chi^2\over 2}\approx
 {\tau_V\over \sqrt{2\pi}}{1\over \gamma 
 \, h/r} \label{eq:ape} \, .
\ee
 Although much simpler than eq.~\refnew{eq:phot}, this form still must be solved numerically.
However, it makes explicit
 the dependencies of $\chi$ on disk properties such as the optical depth and the flaring angle, key physical elements for the instability. 

 In the following, we discuss the main features of this model of a realistic surface.

\medskip

{\bf 1. Equilibrium disk.} \,  
 One can obtain the equilibrium disk  by iteratively solving the  optical surface
equation (eq.~\ref{eq:phot}) and the steady-state thermal equation. Appendix \ref{sec:back} explains how we perform this procedure and achieve convergence.
 The resulting value of $\chi$ for our fiducial disk is shown in Fig.~\ref{fig:timescales}.  It drops only modestly over a large stretch of the disk.
 This can be easily understood from eq.~\refnew{eq:ape}: for $\chi \gg 1$, we can neglect terms outside the exponential to find that $\chi$ depends   logarithmically on $\tau_V$, $\chi \sim  \left(2 \ln\tau_{\rm V} \right)^{1/2}$.
Such a weak dependence on the surface density arises due to the rapid fall-off of density with vertical height, and allows \citetalias{CG97} to assume a constant $\chi$ throughout the disk and to derive a power-law equilibrium solution (eq.~\ref{eq:th2ans}).

\medskip

{\bf 2. Instability.}\, 
The most interesting implication of eq.~(\ref{eq:ape}), however, lies in the relation between $\chi$ and the flaring index $\gamma$. When the flaring  index is larger,  $\chi$ is smaller because the starlight shines closer to the disk's surface normal (larger grazing angle) and so can penetrate deeper towards the midplane. It is this dependence on the grazing angle that is crucial for  the irradiation instability (\S \ref{sec:mech}).

To illustrate this point, we retain the key factors in eq.~\refnew{eq:ape} to recast it as
\be 
e^{-{\chi^2\over 2}}\propto {\dd (h/r)\over \dd  r} \label{eq:simpphot2} \ .
\label{eq:approx3}
\ee
A thermal perturbation  of the spatial form $e^{i k \ln r}$ (with $k \gg 1$) then perturbs the surface as\footnote{\w For a more exact form, c.f. eq.~\refnew{eq:taH}.} 
\be
  \delta \chi \approx - {{ik}\over{\chi \gamma}}{\delta h\over h} \ .
  \label{eq:dchi3}
\ee 
This in  turn affects the heating rate 
as (using eq.~\ref{eq:hthin})
\begin{eqnarray}
 \delta F_{\rm irr} & \propto  & {\dd \over \dd \ln r}\delta H = {\dd \over \dd \ln r}\left(\chi\delta h+h\delta\chi
\right) \nonumber \\
& = & ik\left(
\chi\delta h+h\delta\chi
\right)\, .
\end{eqnarray}
While the first term in the bracket{\w s} is present in the  model of a simplistic surface (constant $\chi$), the second term  is not.
It describes the change in the penetration depth as the grazing angle varies. 
Inserting eq.~\refnew{eq:dchi3} into this expression, we find a positive definite contribution to the heating,\footnote{\w For a more exact form, c.f. eq.~\refnew{eq:dheatapp}.}  
\begin{equation}
\delta F_{\rm irr} \propto i k \chi \delta h + {{k^2}\over{\chi \gamma}} \delta h\, .
\label{eq:newdf}
\end{equation}
 Compared with eq.~\refnew{eq:added}, the second term is new. 
At sufficiently high wavenumbers, it can overcome  damping by radiative cooling  and lead to a new instability, the irradiation instability.

\medskip

{\bf 3.  {\w The Smearing Length}.} \, 
{\w There is a unique scale length associated with irradiation.} Thus far, we   we have assumed that light is  absorbed at a well defined surface (the optical surface),
But in truth, starlight is
deposited over a distance $\Delta r$ (eq.~\ref{eq:deltar}). 
We call the latter the
`smearing  length,' and define an associated wavenumber 
\be
 k_{\rm smear}\equiv {r\over \Delta r}=\chi^2\gamma \ .
 \label{eq:ksmear}
\ee
With this new quantity, eq.~(\ref{eq:newdf}) now reads
\begin{equation}
    \delta F_{\rm irr} \propto  \left(i + {k\over{k_{\rm smear}}}\right)\,k \chi\,  \delta h\,  .
    \label{eq:newdf2}
\end{equation}
For $k\ll k_{\rm smear}$, the simplistic surface model prevails and thermal perturbations lose out to radiative cooling. 
{\w Modes with $k \sim k_{\rm smear}$ or larger, on the other hand, can be destabilized by the changes in the penetration depth of the starlight. 

Moreover, mode growth rates are  affected by the smearing length. 
While the above simple expression indicates that mode growth rate rises with $k$ as $k^2$, a more careful analysis (Appendix \ref{app:ana}) that accounts for a finite smearing length yields a saturated growth rate for modes with $k \gg k_{\rm smear}$ (see text below).
}

\subsection{Linear Perturbations }
\label{sec:linpert}

 After the above qualitative discussions, we now present results from more rigorous derivations. 
 We call the following model `analytical,' to  distinguish it from the RADMC simulations.
 
We employ equations \refnew{eq:thermal0}, \refnew{eq:hthin},
 \refnew{eq:hbyr}, 
 and \refnew{eq:phot} 
to study the  stability  of  a thermal perturbation. We first derive an approximate dispersion relation,  then compare it against exact numerical solutions.  Our analysis shows  that perturbations of certain wavelengths indeed grow in amplitude, on a timescale that is of order the local thermal time.

Assuming a space/time dependence of the form $e^{st+ik\ln r}$ (eq.~\ref{eq:complx}), 
we derive in Appendix \ref{app:ana} an approximate analytic expression for the growth rate,
 \be
s_{\rm grow} \equiv {\rm Re}[s] = 
{1\over t_{\rm th}} { k^2(\chi^2-8)-7k_{\rm smear}^2   
 \over 2( k_{\rm smear}^2+k^2)
 }\, ,
 \label{eq:rgr}
 \ee
where $k_{\rm smear}=\chi^2\gamma$ (eq.~\ref{eq:ksmear}). 
In the limit of long wavelengths ($k\ll k_{\rm smear}$), variations of $\chi$ are relatively insignificant, and so one recovers that the wave damps at the rate $s_{\rm grow}=-7/(2t_{\rm th})$ \citep[eq.~\ref{eq:sold}, and][]{D'Alessio98}. 
 
Equation (\ref{eq:rgr}) allows unstable modes whenever  $\chi > \sqrt{8}$, i.e.,
when the disk is sufficiently opaque  that the optical surface lies well above the gas scale height. As Fig.~\ref{fig:timescales} shows, this  condition  is satisfied throughout our fiducial disk. 
{\w The unstable waves have short wavelengths,
\begin{equation}
 k \geq k_{\rm min} \equiv \sqrt{{7\over{\chi^2-8}}} k_{\rm smear} \, ,
    \label{eq:newk}
\end{equation}
with the value of $k_{\rm min}$ ranging from $4.3$ to $4.8$ in our fiducial
disk (see Fig. \ref{fig:khighlow}).}
{\w For reference, the wavelength of a $k=2.7$ wave spans one decade in radius. So these unstable waves have $\sim 2$ or more wavelengths per radial decade.}

{\w The growth rates of unstable waves first rise with $k$ as $k^2$, before saturating to a constant value for $k \gg k_{\rm smear}$.
This saturation is related to the smearing length, i.e., the finite spread of the optical surface.  When the wavelength is much shorter than this length, changes in stellar heating are subdued relative to that for a  razor-sharp optical surface. As a result, the growth rate saturates.}
 

\begin{figure}[!t]
    \includegraphics[width=1.0\columnwidth,trim=0 150 0 100,clip]{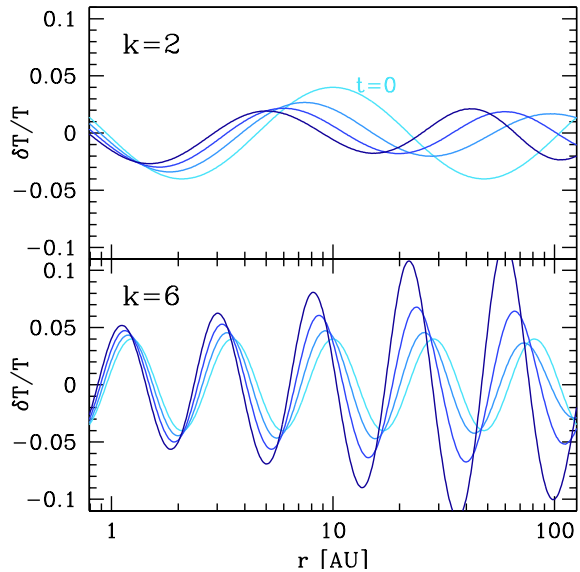}
    \centering
    \caption{
    Evolution of linear perturbations, of the form $e^{i k \ln r}$, in the  model with a  realistic optical surface.  The top panel is for a $k=2$ perturbation, while the bottom is for $k=6$. Snapshots are taken every $1/4$ of the thermal time at 10AU, with the initial ones being the lightest in color.
The $k=2$ perturbation
    decays in time, while the $k=6$ one grows.
    }
    \label{fig:pert_phot}
\end{figure}

\begin{figure}[!t]
    \includegraphics[width=1.0\columnwidth,trim=0 150 0 100,clip]{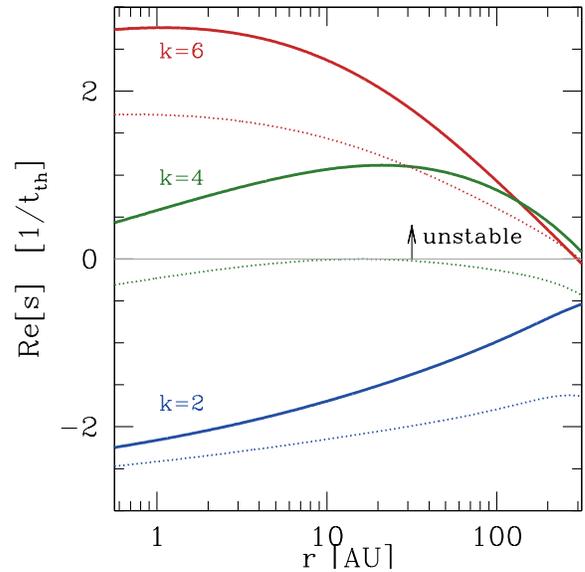}
    \centering
    \caption{Growth rates for linear perturbations of the form $e^{i k \ln r}$, in the model with a realistic surface. 
     The growth rates are in units of inverse local thermal time,  with different colors standing for different $k$-values. The thick curves show the exact  numerical result, while the
    thin dotted curves  present results
    of our analytic approximation
    (eq.~\ref{eq:rgr}).
    {\w Throughout our fiducial disk, $\chi \geq \sqrt{8}$, and there are unstable modes.}
    }
    \label{fig:gr}
\end{figure}

 To confirm these analytical findings, we  obtain mode growth rates by numerically integrating the  relevant equations. The numerical details are in Appendix \ref{sec:evol}. 
Fig.~\ref{fig:pert_phot} shows the results of two such integrations. Starting from  an equilibrium disk (Appendix \ref{sec:back}), we impose  small initial sinusoidal perturbations of the form 
 $e^{ik\ln r}$.
 We observe that the  first perturbation (with $k=2$)  damps with time as the wave travels inward, while the  second one ($k=6$) grows.

We can extract growth rates in integrations like these, following the procedure in Appendix \ref{app:ngr}. The results for some low $k$ waves are plotted in 
Fig.~\ref{fig:gr}.  We find that the analytical expression (eq.~\ref{eq:rgr}) agrees with the numerical results qualitatively. Both suggest that $k \geq 4$ waves should  be unstable over much of our fiducial disk,  with a growth time  that is of order the local thermal time. 


\section{Instability also found by RADMC}
\label{sec:radmc}

We turn to the radiative transfer code RADMC. We do so for multiple purposes. One is to substantiate our analytical results. Another is to use RADMC  to remedy  a  major shortcoming in the analytical approach, i.e., {\w horizontal averaging (see below).} 
  An additional advantage of RADMC is  that it allows us to go beyond the two-frequency approximation.

Given an assumed dust density field, RADMC uses the Monte Carlo method to follow photon  absorption and emission, and  to determine the local equilibrium temperature  by balancing  energy gain and loss. In principle, it can be  combined with the equation of hydrostatic equilibrium to  obtain, iteratively, the  equilibrium disk profile.  This has been attempted by a  number of studies \citep[e.g.,][]{2004A&A...417..159D,2009A&A...497..155M,2012A&A...539A..20S,2019ApJ...871...10U}, either with RADMC or an analogous code. But  the procedure does
 not converge for optically thick disks. We show in Appendix \ref{app:iterative} that  this non-convergence is partly 
caused by a numerical (non-physical) instability. So  non-convergence 
does not prove
that the disk is truly unstable.
To tease out the the physical instability requires some finesse.

\subsection{Simulation Setup}

 We adopt the same fiducial disk as before. 
The density profile is assumed to stay in hydrostatic equilibrium (eq.~\ref{eq:rho}) as  the 
midplane temperature $T = T(r,t)$ evolves. 
RADMC  treats the radiation field at multiple wavelengths.
 The opacity law is set following eq.~\refnew{eq:kap}.

To trace the thermal evolution, 
we replace the stellar heating term in eq.~\refnew{eq:thermal0} by 
\be
{1\over 2} F_{\rm irr}[T] \rightarrow  \sigma_{\rm SB}\left(T_{\rm RADMC}[T]\right)^4\, ,
\label{eq:firr_radmc}
\ee
where $T_{\rm RADMC}[T]$ is the midplane temperature profile output by RADMC.\footnote{{\w We use the midplane temperature, because in RADMC simulations, the disk is vertically isothermal below the optical surface, and the midplane temperature reflects  the amount of stellar heating for the disk. In contrast, the optical surface is hotter \citep{CG97}.}}
We use this as a proxy for stellar heating. The thermal equation becomes
 \be
{3\over 8}\tau_{\rm mm} c_p\Sigma_{\rm  gas} {\dd T\over \dd t} = 
\sigma_{\rm SB}\left( \left(T_{\rm RADMC}[T]\right)^4-T^4\right) \, .
\label{eq:radmc}
\ee
At equilibrium, $T=T_{\rm RADMC}[T]$, as  is desired. Whereas previous works have ignored the left-hand term when  iteratively searching for  equilibrium solution, we integrate  the full equation forward in time. At each time-step, we inject 20 million stellar photon packets from the origin. This provides a sufficiently accurate map of the illumination pattern. We choose a time-step that is a small fraction of the thermal
time (typically 4\% of the thermal time at 10AU). 

 Before we proceed to present results of these integrations, we comment on {\w three} major issues.
{\w The first two cast some doubts on the RADMC results, the third relates to a major improvement of RADMC over our analytical study.}

 First, the inner boundary of our disk is  set to be 1AU.  Between 1 and 1.3AU, we freeze the disk profile to avoid a puffed-up inner rim, which would otherwise  cast a shadow further out.
Although such a shadow may indeed be realistic,  we do not wish its effect to pollute what happens in the simulation domain.
Second,
eq.~(\ref{eq:radmc})  itself contains an important shortcoming:  excessive thermal diffusion.
In realistic situations, temperature gradients in disks should be communicated on diffusive timescales. 
However, the way we hijack RADMC for our purpose effectively assumes that {\w both vertical and} horizontal heat transfer {\w are} instantaneous. {\w The former shortcoming (neglecting vertical diffusion) produces a greater perturbation on the midplane temperature than reality,  boosting the instability; the latter one (horizontal diffusion), on the other hand,}
leads to {\w an} excessive damping of the instability. {\w Clearly, a more thorough study that considers time-dependent radiative transfer is required.}

\begin{figure}[!t]
    \includegraphics[width=1.0\columnwidth,trim=0 0 0 5,clip]{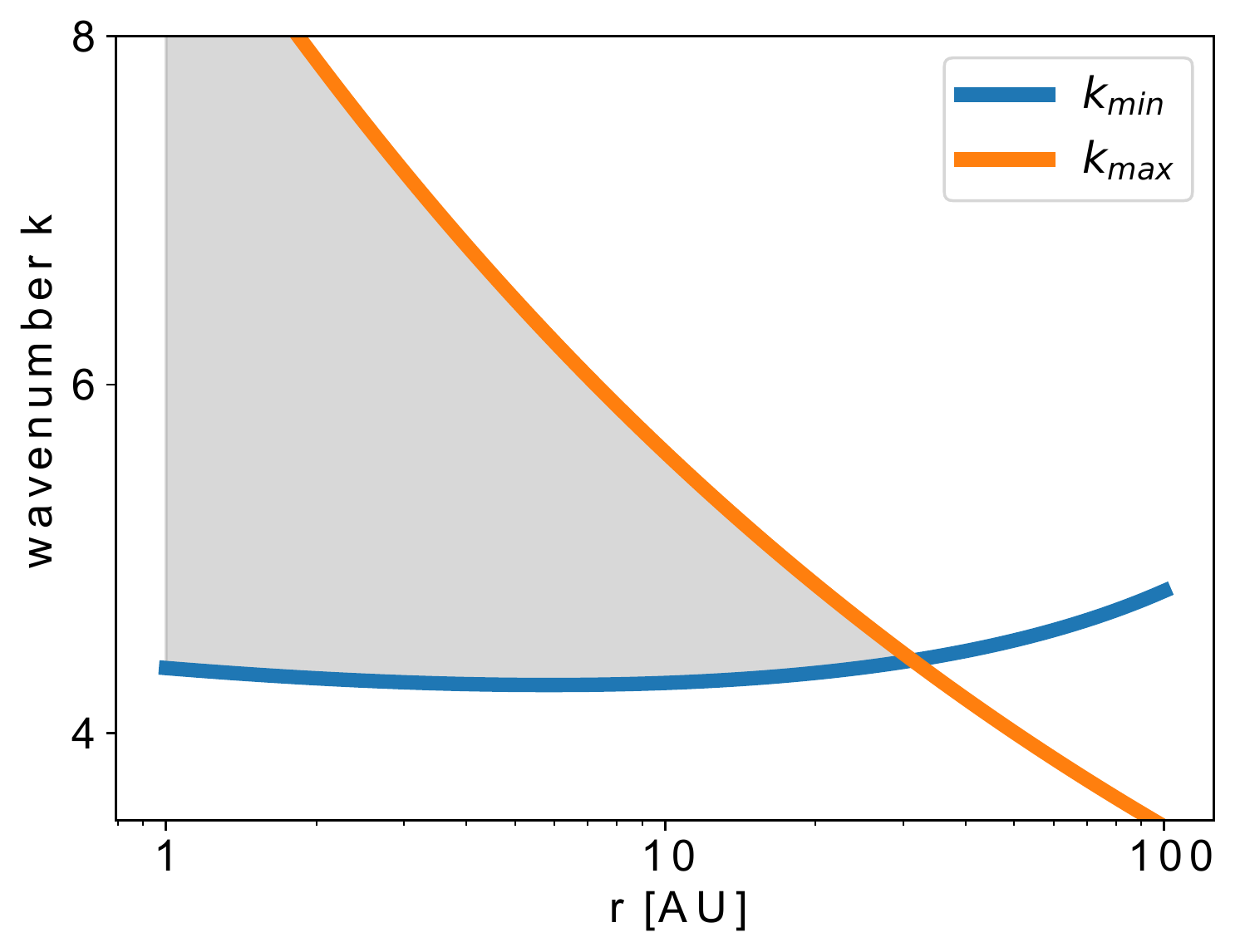}    \centering
    \caption{\w The expected unstable region for our fiducial disk. Wavenumbers of unstable modes have to  fall within the shaded zone: $k_{\rm min} \leq k \leq k_{\rm max}$ (eq.~\ref{eq:newk} \& eq.~\ref{eq:kstable}). This occurs inward of $\sim 30$AU. For reference, $k=2.7$ corresponds to 1 wavelength per radius decade.}
    \label{fig:khighlow}
\end{figure}

{\w The third major issue is what we term 'horizontal averaging'. This is  different from the horizontal diffusion discussed in the last paragraph.}
{\w
As grains at the perturbed optical surface receive more (less) stellar irradiation, they light up as brighter (dimmer) bulbs for the disk down below. In our analytical study, we have assumed that the midplane disk can only see bulbs lying directly above it (heating is only affected by the local gradient, eq.~\ref{eq:hthin}). This assumption fails if the distance between the bulbs and the midplane (more accurately, the vertical photosphere in thermal radiation) is greater than the radial wavelength,  $H \geq r/k$. In this case, the midplane feels a reduced heating (cooling) due to horizontal averaging of the different light bulbs, and we expect the instability to be  quenched. This effect was modelled in \citet{Watanabe}  by manually averaging the perturbations over a radial distance; in RADMC simulations, it is automatically captured.

To order of magnitude, horizontal averaging limits  unstable waves to those with
\begin{equation}
    k \leq k_{\rm max} \equiv {r\over H}\, .
    \label{eq:kstable}
\end{equation}
Together with the lower limit on unstable $k$ (eq.~\ref{eq:newk}), this defines the region in disks where irradiation instability can occur.
For our fiducial disk, we expect unstable waves only inward of $\sim 30$AU (Fig. \ref{fig:khighlow}).} 

\subsection{Linear Evolution}

\begin{figure}[!t]
    \includegraphics[width=1.0\columnwidth,trim=0 100 0 100,clip]{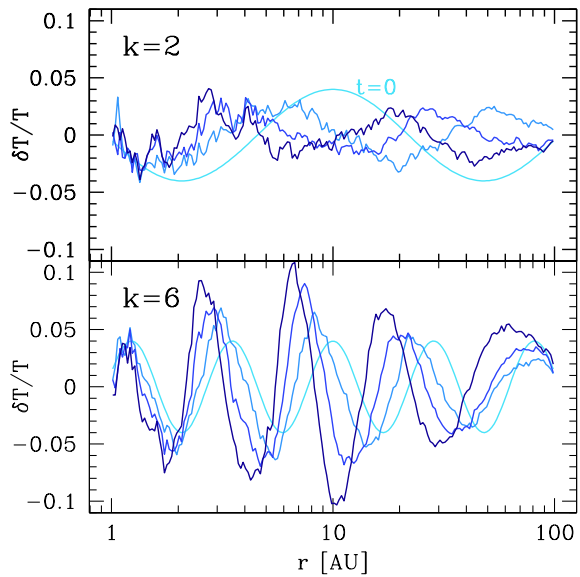}
    \centering
    \caption{ Similar to Fig.~\ref{fig:pert_phot} but now obtained using RADMC integrations. We observe that the  $k=2$ perturbation damps, while the $k=6$ perturbation grows, as is the case in  our analytical model. However, growth for the latter is now restricted to only the inner disk ($r <30$AU).}
\label{fig:rad_pert}
\end{figure}

We impose a  small sinusoidal perturbation of the form $e^{i k \ln r}$, on the equilibrium  
disk (cf. \S \ref{sec:unst}).
 The initial linear evolution is shown in Fig.~\ref{fig:rad_pert}.
Comparing with the analogous plot for our analytical
 model  (Fig.~\ref{fig:pert_phot}), we see that the $k=2$ perturbation decays in  both calculations. On the other hand, while  our analytical model predicts growth for the $k=6$ perturbation at all radii,
RADMC runs show that it only  experiences growth inside 30AU. 
This  reflects the short-coming of the analytical model: it does not account for 
{\w horizontal averaging}, a problem that is more severe in the outer disk. 

\begin{figure}[!t]
    \includegraphics[width=1.0\columnwidth,trim=0 150 0 100,clip]{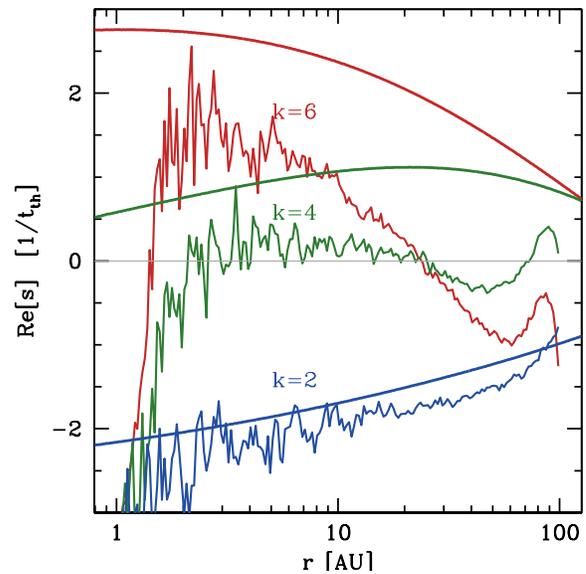}

    \centering
    \caption{
          Growth rates of linear perturbations  extracted from RADMC simulations (jagged thick curves), and
    from the analytical model (smooth thick curves, from Fig.~\ref{fig:gr}). {\w The former show in general lower growth rates -- all modes are stable outside 30AU -- due to both horizontal averaging and horizontal diffusion (see text).}
     RADMC results inward of $\sim 1.3$AU, where we freeze the disk profile, should be ignored. 
    }
    \label{fig:rad_gr}
\end{figure}

We  design a special numerical procedure to extract mode growth rates from RADMC simulations. This is detailed in  Appendix \ref{app:ngr}.
The results are shown in Fig.~\ref{fig:rad_gr} for a few wavenumbers. 
 In this exercise, in order to use RADMC to accurately track  changes in heating associated with a small perturbation, we launch a very large number of photon packets (20 billion). Despite this, the results still appear somewhat jagged.

 We are able to confirm that RADMC yields qualitatively similar growth rates as those from the analytical model (thick lines in Fig.~\ref{fig:gr}, also reproduced here).  But there are two  notable differences. First, RADMC shows that the outer disk (beyond 30AU) is stable to perturbations.  As we explain above, this is related to the issue of  {\w horizontal averaging} in the outer region. Second, the growth rates for unstable modes are  somewhat lower in RADMC,
 indicating  enhanced damping.  
 {\w We believe this may be related to the fact that} the RADMC procedure assumes instantaneous horizontal diffusion. 
A fully time-dependent radiative transfer code, together with a hydrodynamic solver, will be necessary to provide more accurate answers.

\subsection{Nonlinear Development}

\begin{figure}[!t]
    \includegraphics[width=1.0\columnwidth,trim=0 150 0 100,clip]{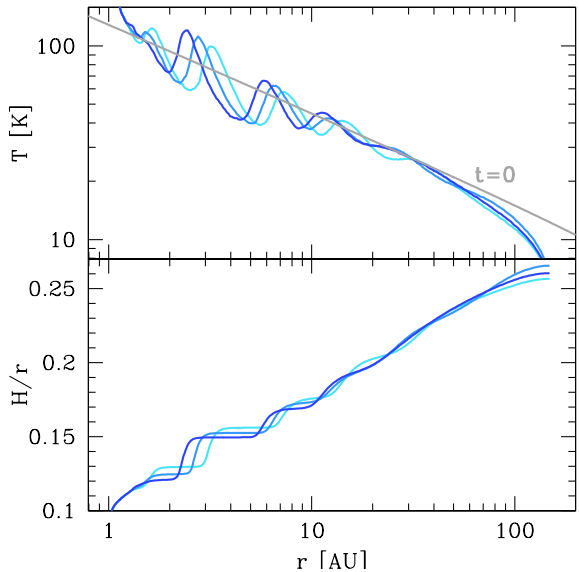}
    \centering
    \caption{Late-time evolution of the midplane temperature (top panel) and the optical surface (bottom panel), obtained from RADMC simulations.
 Starting from  the equilibrium profile (labelled $t=0$), the increasingly dark curves display snapshots at 5, 6, and 7 thermal times at $10$ AU.  The disk inward of 1.3AU is frozen in the simulations. {\w One observes instability only inward of $\sim 30$ AU.}
        }
    \label{fig:temp}
\end{figure}

RADMC also allows us to follow the instability to its nonlinear stage.  For the following runs,  we use 20M photon packets, and have checked to confirm that the behavior does not depend on numerical parameters such as timestep, grid spacing, initial conditions, and inner/outer boundary conditions.  

Starting from an equilibrium profile, 
small perturbations  (seeded only by numerical noise) grow in time. After around 5 thermal times at 10AU, the system reaches a quasi-steady-state, in which waves are generated around $30$AU,  and grow in amplitude as they propagate inwards. 
A few snapshots are presented in Fig.~\ref{fig:temp}. Globally, the most prominent waves have wavenumbers $k\sim $ 4-6  (corresponding to about two wavelengths per radius decade), as expected from  their growth rates
and {\w horizontal averaging (Fig.~\ref{fig:khighlow})} . 
While the disk outside $30$AU remains largely unperturbed, inside $30$AU these waves cause order-unity perturbations in  the midplane temperature. The bottom panel of Fig.~\ref{fig:temp} shows the optical surface
at $0.5\mu$m.
 This looks like a staircase:  the front edge of each stair is illuminated by the star (``stair riser"), and behind that the disk falls into shadow and the optical surface appears flat in $H/r$ (``stair tread").

In the nonlinear stage, the temperature waves are asymmetric with steeper sun-facing edges, as  is also found by \cite{Watanabe}.
This may arise because
at large amplitudes, the local flaring index ($\gamma$)  of the front edge is higher, allowing waves of higher wavenumber to be unstable.\footnote{{\w The inner edge of a wave
could also be thought of as a puffed-up inner rim, casting a long shadow on the disk behind.}}
 As such, perturbations with initially long wavelengths can acquire sharp spatial gradients, leading to  large pressure variations across radial scales as small as the disk scale height. In fact, as our RADMC procedure introduces enhanced horizontal heat diffusion, real disks may harbour waves that are  even larger in amplitudes and sharper in scale.

\section{Re-cap of Technical Results}
\label{sec:summary}

We summarize results obtained in previous sections.

We found that passive disks are susceptible to an  irradiation instability. A slight thermal perturbation in such a disk can lead to the disk receiving even more stellar  heating, thereby initiating unstable growth. We examined this instability using  three
models of increasing complexity and realism. 

In the simplest model,  we assumed that the height of the optical surface (the disk layer at which stellar photons are absorbed) varies in  proportion to the gas scale height  when the disk is thermally perturbed. We found that thermal perturbations propagate inwards and decay as they do so, in agreement with \citet{D'Alessio99}. The modes are damped because the increase in stellar insolation  in this model is  insufficient to counteract the damping due to blackbody cooling.

In contrast, in a model where the optical surface is self-consistently determined, we found that sinusoidal perturbations 
can grow as they propagate inward. The key physics is that as a thermal perturbation
 increases the scale height in a disk annulus, the inner half of the annulus acquires a steeper flaring, and the outer half a more gentle one. Starlight  now shines more overhead for the inner half and so can penetrate into deeper layers; conversely, it is absorbed more shallowly in the outer part. These geometric changes allow the annulus to intercept more starlight, more so than in the first model, and give rise to instability. Thermal perturbations grow and travel inwards in of order the local thermal time. 
 
 In terms of unstable wavelength, very long waves are stable because they cause little change in the surface slope; {\w only waves with wavelengths comparable or shorter}
  than the so-called `smearing length',  the  slant length over which the star deposits its energy, {\w can grow}.
More optically thick disks have shorter smearing lengths and therefore harbour a broader spectrum of unstable waves.
For our fiducial disk, the unstable waves have 
wavenumber $k$ {\w $ \geq 4$, i.e., with $\sim 2$ or more wavelengths per decade in  radius.} 

 To confirm those  
analytical results,
we  retooled the radiative transfer code RADMC
to track the stellar heating.
This also allowed us to address the issue of  {\w horizontal averaging, the complication that arises when the disk heating is not solely determined by the slope of the local optical surface, but also by nearby hot and cold patches on the optical surface. Using both scaling arguments and RADMC simulations, we found  unstable modes exist only inside $\sim 30$AU, or,  where the disk is geometrically thin ($H \leq r/k$). }

We also used RADMC to study the nonlinear evolution of the irradiation instability.
 Within a few thermal times, waves generated far from the star have propagated to the inner region and have grown to order unity amplitudes. The front edges of these waves can be as sharp as a pressure scale height, suggesting  that they may be effective in trapping dust grains. 

\subsection{Relation to 
Watanabe \& Lin}

The irradiation instability was  first discussed in
\citet{Watanabe}. Calculating the optical surface self-consistently (as in our `realistic surface' model), they numerically integrated  the thermal equation and observed that thermal waves are driven to large amplitudes as they travel inwards. But unlike our  RADMC simulations here, they found  instability throughout the whole disk.  This difference may {\w partly be explained by the different ways we  account for  horizontal averaging:
while RADMC naturally captures this phenomenon,  \citet{Watanabe} performed a radial average of the surface re-radiation.}
 Moreover, compared to their numerical results, our analysis here has the advantage of elucidating the origin of the instability, and identifying the unstable wavenumbers.

\section{Examining Assumptions}
\label{sec:assumptions}

There  remain a wide range of uncertainties associated with the irradiation instability.  In this preliminary exploration, we have had to make a string of assumptions. Here, we discuss some of these.\footnote{{\w Also see Appendix \ref{sec:dustgas} which validates the assumption of rapid gas-dust thermal coupling.}} We suggest that modifying them  can lead to an  array of interesting results.  But the essence of the irradiation instability, we feel, should prevail. 

\subsection{Hydrodynamics} 
\label{subsec:hydro}

We simplify the hydrodynamical response of the disk by assuming that the scale height reacts instantaneously to the midplane temperature, the perturbation is axisymmetric, and the surface density does not evolve. These are  questionable for multiple reasons. 

First, they are predicated on the assumption that the orbital time is  shorter
than the thermal time, which is only true inside of $\sim 50$AU in our fiducial
disk  (Fig.~\ref{fig:timescales}). What happens at larger radii is uncertain.  Perturbation analyses by \cite{Dullemond00}
and \cite{2000PhDT........13C} show that a different instability may operates
in that limit. But a more careful analysis, including effects such as starlight penetration depth, smearing length, etc. is needed. In addition, perturbations are likely no longer axisymmetric when the orbital time is long. 

Second, even in the limit of long thermal time, hydrodynamical effects can dramatically alter the story. A local heating will not only change the gas scale height, but can also expel material to neighbouring annuli. This may set up meridional flows that advect heat and dust. 
Vortices or turbulence may also ensue \citep[{\w via, e.g., the Rossby wave instability,}][]{LovelaceLi,Yoram3d}. 
Investigating these effects, both in terms of how they impact the 
irradiation instability and of how they impact the long-term disk evolution,  requires numerical simulations that co-evolve hydrodynamics and radiation. We are currently pursuing this course.
Another  motivation for a full hydro+radiation treatment is to improve on our treatment of thermal diffusion. Thermal perturbations in optically thick disks should propagate as diffusive waves. However, even in our most sophisticated account of the thermal physics (the RADMC approach),  this is not accurately captured. Our retooling of RADMC
accelerates horizontal diffusion artificially. We suspect that this leads to enhanced damping for the waves, as well as reduced amplitudes during the nonlinear evolution.
We have also ignored vertical thermal waves, which may have the potential to weaken the instability  (but see Appendix \ref{subsec:A3}).

\subsection{Dust Movement}
\label{subsec:dust}

\begin{figure}
    \centering
        \includegraphics[width=1.0\columnwidth,trim=0 150 0 100,clip]{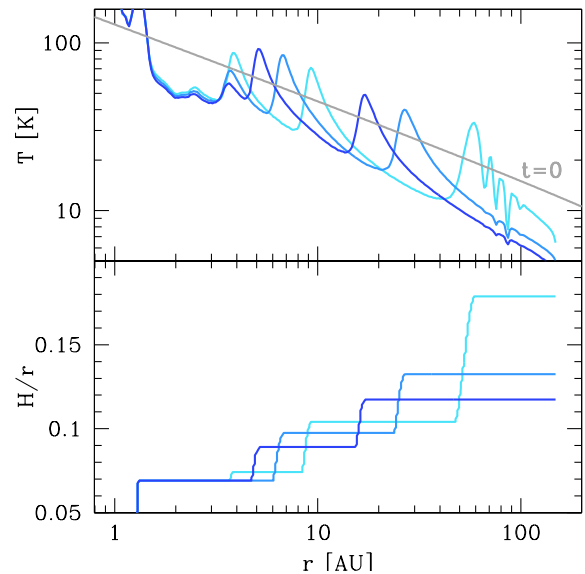}
    \centering
    \caption{An initial foray into dust settling. The set-up is identical to that in Fig.~\ref{fig:temp} except that we imitate dust settling by removing all dust grains above $2$ scale heights. The instability now extends to larger distances (out to $\sim 100$AU).}    \label{fig:settle}
\end{figure}

We assume that dust and gas are co-spatial,  with a constant mass opacity everywhere in the disk.
 This is perhaps our most problematic assumption. It is expected that dust grains evolve in at least three dimensions: they settle vertically, migrate radially, and their sizes can change due to conglomeration and fragmentation. This can change the opacity relative to what we adopt here. 

Dust opacity, especially that in the optical,
is the deciding factor for the irradiation instability. The discussion following eq.~\refnew{eq:rgr} indicates that instability  requires a fairly optically thick disk. 
Opacity in the optical is mostly contributed by micron-sized grains. 
These grains are known to settle quickly from these heights \citep{2004A&A...417..159D}, though even weak turbulence or   circulation can disrupt this process.  While the jury is still out on the vertical distribution of these small grains, we explore a scenario where settling has occurred. In Fig.~\ref{fig:settle}, we repeat our RADMC calculations but now with all dust above $2h$ removed. The optical surface now  
{\w lies much closer to the midplane ($H \leq 2 h$). This mitigates the negative impact of horizontal averaging (eq.~\ref{eq:kstable}).}
{\w And our RADMC simulations show that} the instability now extends to almost the entire disk (out to $\sim 100$AU).

In summary, it appears that a  complete picture of the irradiation instability can only emerge after we understand dust physics, an endeavour that is further complicated by the fact that the hydrodynamical response  of the disk affects dust settling and fragmentation (\S \ref{subsec:hydro}).

\subsection{Inner Region} 
\label{subsec:other}


We  simplified the physics in the inner region substantially, by ignoring the presence of a hot rim and by assuming the star is point-like. {\w Moreover, we have ignored viscous heating, which may dominate over starlight heating in the inner region. Here we briefly discuss how these three effects may impact the irradiation instability.}

 A hot rim is expected to form where the inner disk sees the star unobstructed \citep{Natta01,Dullemond01}. Observed around T Tauri stars \citep{Muzerolle03,Akeson05} and Herbig Ae/Be stars \citep{Natta01},
such a rim casts a long shadow on the disk behind \citep{Dullemond01,Vinkovic06,2012A&A...539A..20S,Flock16}. In this report, we have opted to simplify the picture by freezing the disk profile near the inner edge. But our (additional)
 explorations with RADMC often find that disks near the shadow terminator are unstable, in agreement with some previous work \citep{2012A&A...539A..20S,Flock16,2019ApJ...871...10U}. We suspect that it is related to the irradiation instability, but have yet to provide firm evidence. 

We  also simplified the star into a point-like light source.  This is adequate as long as  $0.4R_*\ll {2\over 7}{H}$ \citep{Kusaka,CG97}, where $R_*$ is the stellar radius, or outside $0.4$AU for a star with a radius of $2.5 R_\odot$. Inside this region, we expect the irradiation instability to be suppressed by the finite-source effect, {\w  and this is indeed demonstrated by the numerical experiments of \citet{Watanabe}.}

{\w Viscous heating associated with mass accretion can compete against stellar heating only in the very inner disk (inward of an AU). The irradiation instability should be suppressed there \citep{Watanabe}.}




\section{Conclusions}
\label{sec:conclusion}

We have developed a model to study the irradiation  instability in passively heated disks.  Our approach, combining analytical and numerical tools, reveals the origin of the instability, identifies the unstable wavelengths and obtains relevant growth-rates.
Our fiducial disk is unstable inside of $30$AU, and produces large-amplitude inwardly-propagating thermal waves.  A preliminary exploration shows that the instability can extend to much further reaches, if some degree of dust settling is included.

Future work should relax many of the approximations we have made here. {\w Of particular urgency are the assumptions of instantaneous hydrostatic adjustment, vertical isothermal structure, and dust evolution.}

If the instability proves robust, it has  many important potential implications.  For example, the large amplitude waves could be the forebears for the  gaps and rings observed by ALMA. {\w These latter features are observed at radial distances of 10-100AU, and have radial scales that are compatible with the unstable 
wavelengths we obtain here ($\sim 2$ wavelengths per decade in radius). The narrowness of the bright rings \citep{Dullemond19} is also consistent with the  nonlinear development of the irradiation instability. Moreover, these waves can affect dust migration and conglomeration, which in turn feeds back on the instability.}
{\w Last but not least,} the same waves could also be hydrodynamically unstable, and produce  vertical circulation, vortices and turbulence. {\w This could provide the much needed viscosity for disk accretion.} 
Clearly, much exciting work lies ahead. 

\acknowledgements

 Y.W. acknowledges support from NSERC.   Y.L. acknowledges NASA grant  NNX14AD21G and NSF grant AST-1352369.  We thank Kees Dullemond for making the RADMC-3D code public and highly usable, {\w and an anonymous referee for a careful and insightful review.}

\bibliographystyle{aasjournal}
\bibliography{gaps}{}

\begin{appendix}

\section{Derivation of  the Thermal Equation}
 \label{sec:thermaleq}

\subsection{Flux Splitting}

In a passively  irradiated disk,  there are three temperatures of concern.  The first  is that at the optical surface, $T_s$. This layer receives the stellar flux $F_{\rm irr}$, then 
 emits  half down, half up. It has negligible thermal inertia and simply acts as a beam splitter. We have
\be
F_{\rm irr} = 2 \sigma_{\rm SB} T_s^4 \, .
\label{eq:a1}
\ee
The downward flux, $\sigma_{\rm SB} T_s^4$, is rapidly reprocessed  at the disk thermal photosphere which has a temperature $T_{\rm ph}$. This layer emits a blackbody flux upwards, and transmits a flux $F_{\rm in}$ downwards toward the optically  thick midplane, leading to 
\begin{equation}
\sigma_{\rm SB} T_s^4 = \sigma_{\rm SB} T_{\rm ph}^4 + F_{\rm in}\, .
\end{equation}
In steady state, $F_{\rm in} = 0$ and the disk is vertically isothermal, with midplane temperature $T_c =  T_{\rm ph} = T_s = (F_{\rm irr}/2\sigma_{\rm SB})^{1/4}$.

\subsection{Simple Thermal Equation}
\label{subsec:A2}

We adopt a simple model for describing the thermal evolution of the disk. We assume the 
 downward diffusive flux determined by the equation of radiative diffusion, \be
F_{\rm in} ={4\sigma_{\rm SB}\over 3} {{ T_{\rm ph}^4 -  T_c^4}\over \tau_{\rm mm}}\, ,
\label{eq:A3}
\ee
where $T_c$ is the midplane temperature and $\tau_{\rm mm}$ is the  optical depth to the midplane  for  thermal radiation.
We then assume that this inward flux heats up gas near the midplane 
at the rate
\begin{equation}
    c_p {\Sigma_{\rm gas}\over 2} \, {{\partial T_c}\over{\partial t}} =  F_{\rm in}\, ,
    \label{eq:A4}
\end{equation}
where $T_c$ is the midplane temperature and $\Sigma_{\rm gas}$/2 is the surface density from the midplane outwards.

{\w Here, we have adopted $c_p$ as the gas heat capacity. \citet{Watanabe1990} have argued for an extra factor of $(\Gamma+1)/2\Gamma$, where $\Gamma$ is the adiabatic index, when one also accounts for the vertical expansion of the disk under stellar gravity.
Moreover, when the disk's thermal state is changed, its rotational speed also has to adapt in order to stay in hydrostatic equilibrium. In this exploration, we have ignored these corrections as they only change the timescales by an order-unity factor.}

Combining  Equations \refnew{eq:a1}--\refnew{eq:A4}, we  land at our desired thermal equation
\begin{equation}
{1\over 2}(1+{3\over 4}\tau_{\rm mm}) c_p \Sigma_{\rm gas}\,  {{\partial T_c}\over{\partial t}} = {1\over 2} F_{\rm irr} - \sigma_{\rm SB} T_c^4\, .
    \label{eq:appthermal}
\end{equation}
In the main text, we drop the 1 in the brackets, as is valid
for $\tau_{\rm mm}\gtrsim 1$.
This derivation ignores heat diffusion in the radial direction,
which is reasonable since the  radial wavelength is  typically much longer than the  scale height.  
It also does not properly account for the vertical temperature
profile, which we discuss next.

\subsection{Vertical Thermal Waves}
\label{subsec:A3}

Equations \refnew{eq:A3}--
\refnew{eq:A4} approximate the vertical transport
of heat when the timescale of variation is longer than the thermal time. Since unstable modes have those two timescales comparable 
to each other, we examine here what happens in the opposite limit.
Our principal conclusions are (i) that vertical thermal waves appear; 
and (ii) 
although those waves could potentially kill
off instability, they probably do not, particularly if the mm-sized
dust has settled.

We solve for the vertical temperature structure when the
 perturbation $\delta T_{\rm ph} \exp(-i \omega t)$. 
Assume the disk is optically thick and energy transport is by radiative diffusion. 
Energy conservation gives  at any height 
\begin{equation} 
\rho_{\rm gas} c_p {{\partial T}\over {\partial t}} = \nabla \cdot \left({{16\sigma_{\rm SB} T^3}\over{3 \kappa \rho_{\rm dust}}} \nabla T\right) \, .
    \label{eq:add1}
\end{equation}
Here, gas provides the thermal capacity ($\rho_{\rm gas}$$c_p$ on the left-hand side), while dust provides the opacity ($\kappa \rho_{\rm dust}$ on the right-hand side).

It is preferable to use optical depth in lieu of the spatial coordinate,  $\tau_z = \int_{z}^\infty \kappa \rho_{\rm dust} dz$. 
Eq.~\refnew{eq:add1} now becomes 
\begin{equation}
{{{3 c_p}\over{16 \sigma_{\rm SB}\kappa}}}{\rho_{\rm gas}\over \rho_{\rm dust}} {{\partial T}\over{\partial t}} = {\partial \over{\partial \tau_z}} \left( T^3 {{\partial T}\over{\partial \tau_z}}\right) \, .
    \label{eq:add4}
\end{equation}
This is valid even if the dust has settled relative to the gas. But  presently we assume no settling. Note that eq.~\refnew{eq:A4} is essentially the  the vertical integral of eq.~\refnew{eq:add4}, together with the approximation that $T^4\propto\tau_z$ {\w in the unperturbed state.}

The solution for the temperature perturbation, subject to the
boundary conditions at the photosphere  and at depth (i.e., $\delta T\rightarrow 0$ at large $\tau_z$) is
\begin{equation}
\delta T (\tau_z, t) = \delta T_{\rm ph} \exp\left(- {{\tau_z}\over{\tau'}}\right) \cos\left(\omega t - {{\tau_z}\over{\tau'}}\right)\, ,
\label{eq:add6}
\end{equation}
where
\be 
\tau' = \left( {32 \sigma_{\rm SB}\kappa T^3\over 3c_p\omega} 
{\rho_{\rm dust}\over\rho_{\rm gas}}\right)^{1/2}=\sqrt{8}{\tau_{z}(z=0)\over \sqrt{\omega t_{\rm th}}}\, .
\ee 
Equation \refnew{eq:add6} describes a wavetrain that propagates
downwards.
The amplitude of the wavetrain decays
exponentially.  The key dimensionless parameter is 
\be 
{\tau_z\over\tau'}\big\vert_{\rm midplane} = {1\over \sqrt{8}}
\sqrt{\omega t_{\rm th}}  \approx \sqrt{{7k\over 32}}= \sqrt{k/4.6} \ , \label{eq:param}
\ee 
where in the approximate equality we used the dispersion relation for irradiated thermal waves  (eq.~\ref{eq:sold}).
If that parameter is $\gtrsim {\pi\over 2}$, then the 
vertical temperature profile will be oscillatory, which will
likely suppress  the irradiation instability. 
Our unstable modes have $k\sim 6$, which suggests that
the instability is not suppressed.
But clearly
a more careful treatment
is needed to obtain more accurate values for the order-unity
coefficents.

One should also account for dust settling, because the mm-sized grains responsible for the
opacity tend to settle towards the midplane. 
The waves in eq.~\refnew{eq:add6} are in $\tau_z$ rather than $z$. 
So if the grains  settle into a thin layer, then even if the parameter in eq.~\refnew{eq:param} exceeds unity, 
the temperature throughout most of the lowest gas scale height (by volume)
will be similar to that at the photosphere. 
Thus dust settling will likely help prevent thermal waves
from suppressing  instability.

\section{Methods for the Model with Realistic Surface}

\subsection{Time Integration}
\label{sec:evol}

 The model equations of motion are listed  in \S \ref{sec:linpert}.  
We evolve the thermal equation forward in time with the Euler method.
The spatial derivative
in equation (\ref{eq:hthin}) is,  at the $i$th
gridpoint,   taken to be proportional to $H/r\vert_{i+1}-H/r\vert_i$---i.e.,, 
the first-order upwind scheme. 
The  grid in $r$ is logarithmic, with typically 250 gridpoints
per decade. 
 At each timestep, 
 $H(r)$ on the spatial grid is obtained from $T(r)$ as follows:
after converting from $T$ to $h$ with the hydrostatic equation, 
the integral in the  equation for the optical surface (eq.~\ref{eq:phot}) is performed numerically at the $i$th gridpoint for an assumed
value of $H(r_i)$; and that $H(r_i)$ is then adjusted via a root-finder
until the integral is   unity. 
For the boundary conditions, we typically freeze the temperature near the inner and outer
boundary.

\subsection{Solving for Equilibrium}
\label{sec:back}

As stated in \S \ref{sec:phys}, we obtain the equilibrium disk profile by
iteratively solving the steady state thermal equation 
(eq.~\ref{eq:bg}) and  the photosphere equation (eq.~\ref{eq:phot}). 
A naive application of this procedure is typically numerically unstable. But that instability may be avoided by using the integral of the
 thermal equation in place of the thermal equation itself. 
In particular,   for any assumed  $\chi(r)$ profile one may integrate equation \refnew{eq:bg} to yield 
\be
{H\over r} = 0.02 \, \chi_{\rm eff}^{8/7}\, r_{\rm AU}^{2/7}
\label{eq:HR}
\ee
after using hydrostatic equilibrium and defining  $\chi_{\rm eff}$ via
 ${1\over r^2\chi_{\rm eff}^8}\equiv \int_r^\infty{2d\ln r'\over r'^2\chi(r')^8}$.
Therefore the iteration proceeds as follows:  (i) given a profile for $\chi(r)$
(initially taken to be constant), calculate $H$ via equation (\ref{eq:HR}), and
thence $h=H/\chi$; 
(ii) use that $h$ in the equation for the optical surface, and solve for $H(r)$, and thence
 $\chi=H/h$; (iii) insert that $\chi$ back into step (i) and repeat until convergence.

\subsection{Measuring Linear Growth Rates Numerically}
\label{app:ngr}

We explain how we measure the exact growth rates 
shown in Figs. \ref{fig:gr} and \ref{fig:rad_gr}. 
It is non-trivial because the growth rate is in general complex-valued. 
We start
from the linearized thermal equation (eq.~\ref{eq:lin}), which is valid for arbitrary forms
of $F_{\rm irr}[T]$, and
 take the perturbed temperature to be of the form 
${\delta T\over T}={\rm Real}(\bld{\epsilon_T}e^{ik\ln r})$, 
where in this appendix
we denote complex numbers in bold.  
As a result,
${\delta F_{\rm irr}\over F_{\rm irr}}={\rm Real}(\bld{\sigma \epsilon_T}e^{ik\ln r})$, where 
$\bld{\sigma}$ is a proportionality constant whose value we determine
as follows.
First, we write $\bld{\sigma}$ and $\bld{\epsilon_T}$
in amplitude-phase form: $\bld{\sigma}=\sigma e^{i\phi_\sigma}$ and $\bld{\epsilon_T}=\epsilon_Te^{i\phi_T}$, which implies
that
\be
{\delta F_{\rm irr}\over F_{\rm irr}}=\sigma\epsilon_T \cos(\phi_\sigma+\phi_T+k\ln r) \ .
\label{eq:dgam1}
\ee
And second,
 we evaluate the perturbed $F_{\rm irr}$ numerically, via
\be
{\delta F_{\rm irr}\over F_{\rm irr}}
={ {F_{\rm irr}}[T\left(1+{\rm Real}(\bld{\epsilon_T}e^{ik\ln r})\right)]
\over F_{\rm irr}[T]} -1\, . \label{eq:dgam}
\ee
To
evaluate $\bld{\sigma}$ at any desired $r$ and $k$, we 
  choose a fixed value of  $\epsilon_T\ll 1$ and various values of $\phi_T$, and
   evaluate Equation (\ref{eq:dgam}) at those values.  
  The resulting function of $\delta F_{\rm irr}/F_{\rm irr}$   versus $\phi_T$
  is fit to a sinusoid with the form of equation \refnew{eq:dgam1}, which allows
  us to extract the values of $\sigma$ and $\phi_\sigma$--and hence $\bld{\sigma}$. 
  
  With the value of $\bld{\sigma}$ in hand, we substitute into equation \refnew{eq:lin} the relation between
  the complex amplitudes, 
  ${\bld{\delta F_{\rm irr}}\over F_{\rm irr}}=\bld{\sigma}\bld{\epsilon_T}$.  
  Equation  (\ref{eq:lin}) then yields the solution $\bld{\epsilon_T}\propto e^{\bld{s} t}$, where
  the (complex) growth rate is
  \be
   \bld{s} = {1\over t_{\rm th}}\left( \bld{\sigma}-4\right)\, .
  \ee

\subsection{Analytical  Growth Rate}
\label{app:ana}
 
We derive the growth rate analytically, under a number of simplifying assumptions. 
To begin, we first determine the relationship between the perturbed
temperature and the perturbed heating rate. 
When the temperature is perturbed by $\delta T$,  hydrostatic equilibrium 
provides the perturbation in scale height ($\delta h$), and the  equation
for the optical surface
(eq.~\ref{eq:phot}) then yields $\delta H$, which in turn leads to the perturbed heating rate.

Under perturbations $\delta h(r)$ and $\delta H(r)$ (but no change in surface density), the perturbed equation for the optical surface (eq.~\ref{eq:phot}) is Taylor-expanded to  linear order, which yields
\be
0 = \int_0^r {dr'\over r'}{2\tau_V(r')\over\sqrt{2\pi}\left({h(r')\over r'}\right)}e^{- {1\over 2}{H^2\over r^2}{r'^2\over  h(r')^2}}
\times \left\{  {{H^2}\over {r^2}} \, {{r'^2}\over{h(r')^2}}\,  \left[ - {{\delta H}\over H} + {{\delta h(r')}\over{h(r')}}\right]
- {{\delta h(r')}\over{h(r')}}\right\}\, ,
\label{eq:square}
\ee
where $H$ and $\delta H$ are understood to be functions of $r$. In the limit that $\chi = H/h \gg 1$, 
the last term inside the curly brackets is small and we obtain 
\be
{\delta H\over H} =
{\rm NUM\over DEN}
\ee
where the numerator and the denominator are, respectively,
\be
{\rm NUM}=  \int_0^r  {dr'\over r'}{2\tau_V(r')\over\sqrt{2\pi}\left({h(r')\over r'}\right)^3}{\delta h(r')\over h(r')}
e^{- {1\over 2}{H^2\over r^2}{r'^2\over  h(r')^2}}\, , \hskip0.5in
{\rm DEN} = 
 \int_0^r  {dr'\over r'} {2\tau_V(r')\over\sqrt{2\pi}\left({h(r')\over r'}\right)^3}e^{- {1\over 2}{H^2\over r^2}{r'^2\over  h(r')^2}}
\ee
One may use a trick to perform the DEN integral: 
{\w taking the derivative of the equation for the optical surface (eq. \ref{eq:phot}), one finds}
\be 
0 &=& {d\over dr}\int_0^r  dr'{2\tau_V(r')\over\sqrt{2\pi}h(r')}e^{- {1\over 2}{H(r)^2\over r^2}{r'^2\over  h(r')^2}} \\
&=&{2\tau_V(r)\over\sqrt{2\pi}h(r)}e^{- {\chi(r)^2\over 2}}
+\int_0^r  dr'{2\tau_V(r')\over\sqrt{2\pi}h(r')}e^{- {1\over 2}{H(r)^2\over r^2}{r'^2\over  h(r')^2}}\left(-{H(r)\over r}{d(H/r)\over dr}{r'^2\over h(r')^2}  \right) \\
&=&{2\tau_V(r)\over\sqrt{2\pi}h(r)}e^{- {\chi(r)^2\over 2}}
-\left({H\over r}{dH/r\over dr}\right)\times {\rm DEN}
\ee
{\w where $\chi(r)=H(r)/h(r)$. The latter equations may be solved for DEN to yield}
\be
{\rm DEN} ={1\over {H\over r}{d{H\over r}\over d\ln r}} {2\tau_V\over\sqrt{2\pi}(h/r)}e^{-{\chi(r)^2\over 2}} \approx {1\over{\chi^2 \gamma}}\, {2\tau_V\over\sqrt{2\pi}(h/r)^3}e^{-{\chi^2\over 2}} \, ,
\ee 
{\w after approximating}
$d \ln {H/r}/d\ln r \approx   \gamma$, with
$\gamma$ being the logarithmic slope of the background $h/r$ (eq.~\ref{eq:gammadef}), as is valid for large enough $\chi$.

To perform the NUM integral, we write the temperature perturbation to be in the complex form  of eq.~\refnew{eq:complx}.
We also have
${\delta h\over h}={1\over 2}{\delta T\over T}$ in hydrostatic equilibrium.
We may now approximate the  NUM integral as we did for the background approximate 
equation for the optical surface (see footnote \ref{foot:garaudapprox}) to obtain
\be
{\rm NUM} \approx {1\over 2}
{1\over ik +\chi^2\gamma} 
 {2\tau_{\rm V}\over\sqrt{2\pi}(h/r)^3 }
e^{-{\chi^2\over 2}}\, 
{{\delta T}\over T}\, 
\,
   ,
\ee
Combining the above two results, we arrive at 
\be
{\delta H\over  H} =
 {1\over 2}
 {\chi^2\gamma \over  ik+\chi^2\gamma}  {\delta T\over T} 
 \label{eq:taH}
\ee
 For comparison, the simplistic surface model (constant $\chi$) gives $\delta H/H = \delta h/h = {1\over 2} \delta T/T$, which agrees at small $k$.

We may now insert $\delta H/H$ into 
the heating rate (eq.~\ref{eq:hthin}), which yields
\be
{\delta F_{\rm irr}\over F_{\rm irr}}={{d \over d \ln r}\left({\delta H\over H}{H\over r}\right)\over {d\over d\ln r}{H\over  r}}
= \left({ik\over\gamma}+1\right) {\delta H\over H}=
\left({\chi^2\over 2}{ik+\gamma \over  ik+\chi^2\gamma}\right) \, {{\delta T}\over T}\ ,
 \label{eq:dheatapp}
\ee
where in the second equality we assume that the spatial variation of Eq.~\refnew{eq:taH} is 
dominated by the sinusoidal dependence, as is reasonable for $k\gtrsim $ a few.
 Inserting this 
 into the linearized thermal equation (eq.~\ref{eq:lin})  yields  the complex growth rate 
\be
s \approx {1\over 2t_{\rm th}}\left[{ik(\chi^2-8)-7\chi^2\gamma\over ik+\chi^2\gamma}  
\right] \ .
\label{eq:gr}
\ee
\label{sec:yyy}

\section{Numerical  Instability in  RADMC due to Iterations}
\label{app:iterative}

As described in  \S \ref{sec:radmc}, a number of papers attempt to find the equilibrium state of passive disks  with an iterative scheme, based on a radiative transfer code such as RADMC.  These often find that the iterations do not converge,   but instead produce large-amplitude propagating waves \citep[e.g.,][]{2004A&A...417..159D,2009A&A...497..155M,2012A&A...539A..20S,2019ApJ...871...10U}. Here we show that the reason for the non-convergence is an 
instability closely connected with the irradiation instability described in this paper. 
However, the two instabilities are not identical:  the iteration instability is partly
polluted by a numerical (i.e., non physical) instability, and so if the disk 
is unstable under iterations, it  is not necessarily unstable in reality.

As described in \S \ref{app:ngr}, the key quantity governing the stability of a disk
is the complex number $\bld{\sigma}$ that relates heating perturbations to temperature perturbations,  defined via
\be 
{\bld{\delta {F_{\rm irr}}}\over F_{\rm irr}}=\bld{\sigma}{\bld{\delta T}\over T} \ .
\ee 
In the above, the bold $\bld{\delta T}$ is a complex amplitude, i.e., the real 
temperature perturbation is $\delta T={\rm Real}\left(\bld{\delta T}e^{ik\ln r}  \right)$, 
and similarly for $\bld{\delta F_{\rm irr}}$; $\bld{\sigma}$ is a function of both 
$r$ and $k$, and its value may be determined numerically (\S \ref{app:ngr}). 
Given the value for $\bld{\sigma}$, the linearized thermal equation (eq.~\ref{eq:lin}) shows that the 
disk is unstable if ${\rm Real}(\bld{\sigma})>4$;  otherwise, it is stable.

Now, let us compare this behavior with that of the iteration scheme, 
in which the temperature at the $k$th iterative step ($T_k$) is
the equilibrium solution of the thermal equation, i.e., 
from eq.~\refnew{eq:radmc},
$T_k=T_{\rm RADMC}[T_{k-1}]$.
Equivalently, writing this in terms
of $F_{\rm irr}$ (eq.~\ref{eq:firr_radmc}),
\be 
T_k = \left( {F_{\rm irr}[T_{k-1}]/2\over \sigma_{\rm SB}}  \right)^{1/4}
\label{eq:tk}
\ee 
Setting $T_k=T+\delta T_k$, where $T$ is the equilibrium temperature, 
and linearizing eq.~(\ref{eq:tk}) yields
\be 
{\bld{\delta T}_k\over T}={1\over 4}{\bld{\delta F_{\rm irr}}_{,k-1}\over F_{\rm irr}}={\bld{\sigma}\over 4}
{\bld{\delta T}_{k-1}\over T}
\ee 
The solution to this difference equation is $\bld{\delta T}_k={\rm const}\times \left( {\bld{\sigma}\over 4}  \right)^k$.  Therefore the solution is unstable if 
$|\bld{\sigma}|>4$, which is less stringent than the true criterion for
instability (${\rm Real}(\bld{\sigma})>4$). 
We note that typically the real and imaginary parts of $\bld{\sigma}$ are comparable
to each other, and so the criterion for iterative instabiility is incorrect by 
an order-unity factor. For example, in the   model with a realistic surface, the
marginal $k$ for stability (plotted in Fig.~\ref{fig:gr}) is $\sim$2 for the
iterative instability---rather than the true value of $\sim 3-4$ as is shown in the figure. 

\section{Gas-Dust Thermal Coupling}
\label{sec:dustgas}

{\w In our work, we have assumed that the gas and dust components in the disk share the same temperature. 
This requires efficient energy transfer.}
{\w We examine the validity of this assumption here.

We first focus on the  midplane gas.
As gas is inert radiatively, it only receives or loses energy when colliding with the dust grains.  Assuming all dust particles have size $s$ and bulk density $\rho_{\rm bulk}$, we can estimate the timescale for changing the gas's thermal energy as,
\begin{equation}
\tau_{\rm gas-dust} \sim {{s\rho_{\rm bulk}}\over{c_s \rho_{\rm dust}}}\, .
\label{eq:gasdustrate}
\end{equation}
Comparing this to the dynamical time, and evaluating at the midplane of our adopted disk (eq.~\ref{eq:sigmadust}), we find
\begin{equation}
    \Omega \, \tau_{\rm gas-dust} \sim {{s \rho_{\rm bulk}}\over{\Sigma_{\rm dust}}} \sim 0.005 \, \left({s\over{1\rm mm}}\right) \left({{\rho_{\rm bulk}}\over{1 {\rm g/cm^3}}}\right) r_{\rm AU}\, .
    \label{eq:gasdustrate2}
\end{equation}
This process is therefore fast throughout the disk midplane. The strong coupling between gas and dust may fail at high altitudes. Fortunately, this should not affect the irradiation instability.
}

\end{appendix}

\end{document}